\documentclass[%
 reprint,
superscriptaddress,
showkeys,
twocolumn,
amsmath,amssymb,
aps,
prf,
]{revtex4-2}

\usepackage{graphicx}
\usepackage{dcolumn}
\usepackage{bm}
\usepackage{soul}
\usepackage{hyperref}

\usepackage{amsfonts}
\usepackage{amssymb}
\usepackage{bbold}
\usepackage{booktabs}
\usepackage{ulem}
\usepackage{xr}
\usepackage{tikz}
\usetikzlibrary{shapes}

\usepackage{soul}
\newcommand{\Ca}{\text{Ca}}
\newcommand{\Rey}{\text{Re}}
\newcommand{\Uw}{\text{U}_{\mbox{\scriptsize w}}}
\newcommand{\Bq}{\text{Bq}}
\newcommand{\Bqs}{\text{Bq}_{\mbox{\scriptsize s}}}
\newcommand{\Bqd}{\text{Bq}_{\mbox{\scriptsize d}}}
\newcommand{\muin}{\mu_{\mbox{\scriptsize in}}}
\newcommand{\muout}{\mu_{\mbox{\scriptsize out}}}
\newcommand{\mum}{\mu_{\mbox{\scriptsize m}}}
\newcommand{\viscratio}{\tilde{\mu}}

\newcommand{\effi}{f_i}

\newcommand{\Kal}{k_{\scriptsize \alpha}}

\newcommand{\Nf}{N_{\mbox{\scriptsize f}}}

\newcommand{\kB}{k_{\mbox{\scriptsize B}}}
\newcommand{\kS}{k_{\mbox{\scriptsize S}}}

\newcommand{\WB}{W_{\mbox{\scriptsize B}}}
\newcommand{\WS}{W_{\mbox{\scriptsize S}}}

\newcommand{\cs}{c_{\mbox{\scriptsize s}}}
\newcommand{\mus}{\mu_{\mbox{\scriptsize s}}}
\newcommand{\mud}{\mu_{\mbox{\scriptsize d}}}
\newcommand{\tc}{t_{\mbox{\tiny c}}}

\newcommand{\dA}{d_{\mbox{\scriptsize A}}}
\newcommand{\dT}{d_{\mbox{\scriptsize T}}}

\newcommand{\Lx}{L_{\mbox {\tiny x}}}
\newcommand{\Ly}{L_{\mbox {\tiny y}}}
\newcommand{\Lz}{L_{\mbox {\tiny z}}}

\newcommand{\lambdain}{\lambda_{\mbox {\tiny in}}}
\newcommand{\lambdafin}{\lambda_{\mbox {\tiny fin}}}

\newcommand{\ml}{\sigma}

\newcommand{\dotgammasmall}{\dot{\gamma}_{\mbox{\tiny small}}}
\newcommand{\fsmall}{F_{\mbox{\tiny small}}}

\newcommand{\sigmat}{\sigma^{\mbox{\tiny T}}}
\newcommand{\sigmaf}{\sigma^{\mbox{\tiny F}}}
\newcommand{\sigmap}{\sigma^{\mbox{\tiny p}}}

\renewcommand{\vec}[1]{\ensuremath{\boldsymbol{#1}}}

\begin{document}
\preprint{APS/123-QED}
\title[Sample title]{
On the effects of membrane viscosity on transient red blood cell dynamics}

\author{Fabio Guglietta}
\email{fabio.guglietta@roma2.infn.it}
\affiliation{Department of Physics \& INFN, University of Rome ``Tor Vergata'', Via della Ricerca Scientifica 1, 00133, Rome, Italy}
\affiliation{Chair for Computational Analysis of Technical Systems (CATS),\\ {\it RWTH Aachen University}, 52056 Aachen, Germany.}
\affiliation{Computation-based  Science  and  Technology  Research  Center, {\it The  Cyprus  Institute},\\ 20 Konstantinou Kavafi Str., 2121 Nicosia, Cyprus.}
\author{Marek Behr}
\affiliation{Chair for Computational Analysis of Technical Systems (CATS),\\ {\it RWTH Aachen University}, 52056 Aachen, Germany.}
\author{Luca Biferale}
\affiliation{Department of Physics \& INFN, University of Rome ``Tor Vergata'', Via della Ricerca Scientifica 1, 00133, Rome, Italy}
\author{Giacomo Falcucci}
\affiliation{Department of Enterprise Engineering “Mario Lucertini,” {\it University of Rome ``Tor Vergata"},\\ Via del Politecnico 1, 00133 Rome, Italy.}
\affiliation{John A. Paulson School of Engineering and Applied Physics, {\it Harvard University},\\  33 Oxford Street, 02138 Cambridge, Massachusetts, USA.}
\author{Mauro Sbragaglia}
\affiliation{Department of Physics \& INFN, University of Rome ``Tor Vergata'', Via della Ricerca Scientifica 1, 00133, Rome, Italy}

\begin{abstract}
Computational Fluid Dynamics (CFD) is currently used to design and improve the hydraulic properties of biomedical devices, wherein the large scale blood circulation needs to be simulated by accounting for the mechanical response of red blood cells (RBCs) at mesoscales. In many practical instances, biomedical devices work on time-scales comparable to the intrinsic relaxation time of RBCs: thus, a systematic understanding of the {\it time-dependent} response of erythrocyte membranes is crucial for the effective design of such devices. So far, this information has been deduced from experimental data, which do not necessarily adapt to the broad variety of the fluid dynamic conditions that can be encountered in practice. This work explores the novel possibility of studying the time-dependent response of an erythrocyte membrane to external mechanical loads via mesoscale numerical simulations, with a primary focus on the detailed characterisation of the RBC relaxation time $\tc$ following the arrest of the external mechanical load. The adopted mesoscale model exploits a hybrid Immersed Boundary-Lattice Boltzmann Method (IB-LBM), coupled with the Standard Linear Solid model (SLS) to account for the RBC membrane viscosity. We underscore the key importance of the 2D membrane viscosity $\mum$ to correctly reproduce the relaxation time of the RBC membrane. A detailed assessment of the dependencies on the typology and strength of the applied mechanical loads is also provided. Overall, our findings open interesting future perspectives for the study of the non-linear response of RBCs immersed in time-dependent strain fields.
\end{abstract}

\maketitle
\section{Introduction}\label{sec:intro}
Blood  is a  heterogeneous non-Newtonian fluid, mainly composed by plasma (a Newtonian fluid which constitutes approximately the $55\%$ of the total volume) and red blood cells (RBCs) (the hematocrit, i.e. the concentration of erythrocytes within the blood, ranges between $37\%$ to $50\%$, according to age and sex)~\cite{popel2005microcirculation,baskurt2007handbook,fung2013biomechanics,thesis:kruger,thesis:mountrakis,falcucci2019simulating}. 
On one hand, the information on the mechanical and biochemical responses of RBC membrane can give useful information on human health state~\cite{mills2004nonlinear}. On the other hand, it can be used to improve the design of biomedical equipment, such as Ventricular Assist Devices (VADs), which have been developed in the years to assist blood circulation in patients with heart failures~\cite{art:behbahani09,murakami1979nonpulsatile,nonaka2001development}. The optimal design of VADs requires a thorough understanding of the fluid dynamic field inside these devices, something not realisable from experimental measurements due to the difficulty to have non intrusive probes dispersed in the fluid volume. Computational Fluid Dynamics (CFD) offers a solution to this problem, and it has been extensively used to improve VADs hydraulic properties, through the simulation of their internal blood circulation, and providing an augmented reality representation of the flow field inside the whole device ~\cite{art:anderson00,art:miyazoe98,art:yu00,art:qian00,art:allaire99,art:burgreen01}.\\    
One of the most challenging aspects in designing VADs is connected to the reduction of hemolysis, \cite{art:behbahani09,art:arora04}, that is the release of the cytoplasm (i.e. the fluid inside the membrane of the RBC). Hemolysis is directly connected to the fluid-dynamical stress experienced by RBC membrane~\cite{art:arora04,thesis:pauli}. The corresponding CFD framework is rather challenging, due to the intrinsic {\it multiscale} nature of the involved physics, with erythrocytes (characterised by an average diameter of $d\sim 9\times 10^{-6} \mbox{ m}$, at rest) evolving within the VAD, whose internal chambers are in the order of $\sim 0.01\ \mbox{m}$. 
Many medical issues require a description over several order of magnitude both in time and in space; multiscale approaches for cardiovascular applications are growing in the last years, and they have been adopted to address different situations~\cite{owen2018structural,quarteroni2016geometric}: for example, they have been used to investigate the growth of the abdominal
aortic aneurysm~\cite{watton2004mathematical}, or to study the properties of the myocardium~\cite{owen2018structural}.
In addition, RBCs dynamics within the VAD is {\it non-steady}, since their residence time through the device is comparable to their membrane (intrinsic) relaxation time, $\sim 0.500\ \mbox{s}$~\cite{art:arora04}. To account for these complex characteristics, Arora {\it et al.}~\cite{art:arora04} developed a ``tensor-based'' model (also known as strain-based model), in which the time-dependent shape of a single RBC is described by a differential equation that accounts for a few mesoscale parameters. Such a slender model is, then, coupled to the Navier-Stokes hydrodynamics within the VADs~\cite{art:arora2006hemolysis,thesis:pauli,art:pauli2013transient} through Finite Element (FEM) simulations.  Hemolysis is then quantified in terms of the ratio between plasma free haemoglobin to total haemoglobin (i.e. $\Delta Hb / Hb$), which depends on the effective shear rate. Such a tensor-based approach allows for a convenient and efficient multiscale simulation: given few ``mesoscale'', non-steady parameters in input (such as the membrane relaxation time $\tc$, the degree of time-dependent deformation under flow), a ``large scale'' numerical simulation of the pump can be carried out, with the aim of designing a suitable geometry that minimises hemolysis~\cite{art:arora2006hemolysis,art:pauli2013transient,art:gesenhues2016strain}. 
In such models, the RBC is described via a reduced model comprising the competing actions of elastic forces, which tend to relax the cell to its original shape, and drag forces, which tend to deform the RBC. Understanding the precise dependency of the relaxation time on the degree of initial deformation for realistic values of membrane viscosity is definitively important for novel applications in the large scale numerical simulation of blood flows within VADs.
So far, the free mesoscale parameters needed by Arora's model have been recovered from experimental data~\cite{art:henon99}, which do not obviously adapt to the the broad variety of conditions that can be encountered in practice. In particular, some delicate issues pertain the relaxation time $\tc$ of RBCs after the arrest of external mechanical loads: viscous effects play a key role and come both from the internal (cytoplasm) and external (plasma) fluid, and from the membrane itself.
Pioneering experimental studies focused on the relaxation of RBCs in a variety of experimental set-ups~\cite{art:henon99,art:hochmuth79,art:baskurt96,art:bronkhorst95}, while more recent experiments target the shape deformation time in microfluidic set-ups~\cite{art:tomaiuolo11}. In these experiments, membrane viscosity $\mum$ has shown a direct role in determining the relaxation time of the whole RBC~\cite{art:prado15,art:tran84,art:tomaiuolo11}, with old erythrocytes characterised by a higher membrane viscosity compared to younger cells~\cite{art:tran84,art:henon99}. The membrane viscosity is not known \textit{a priori} and is usually computed indirectly from $\tc$, with larger membrane viscosities generically associated to larger relaxation times. For an accurate determination, various phenomenological models have been proposed in the literature~\cite{evans1976membrane,art:prado15}, resulting in consistent discrepancies in the values of $\mum$.\\ 
In this heterogeneous panorama, the comprehensive characterisation of the RBC relaxation time at the sudden arrest of an external mechanical load considering different membrane viscosities is still lacking. Experimental data, in fact, do not allow a comprehensive characterisation of this phenomenon, due to the intrinsic difficulties in isolating the effects of membrane properties and external/internal flow properties. Rather, mesoscale simulations represent a more appropriate tool, wherein one can easily modify physical properties such as membrane characteristics (elasticity, viscosity), or focus on the specific flow conditions that trigger membrane peculiar responses. It must be remarked, however, that while mesoscale simulations of RBCs have been carried out in a variety of contexts (for example, to investigate the rheology of RBCs suspensions~\cite{thesis:kruger,thesis:mountrakis,thesis:janoschek,gross2014rheology}, the tumbling-to-tank treading transition~\cite{art:kruger13}, the deformation of RBCs~\cite{art:fedosov10,art:pan11,art:fedosov2010systematic,art:kruger14deformability}, the dependence of blood viscosity on shear rate and hematocrit~\cite{art:fedosov2011predicting}, the shape phase diagram~\cite{art:guckenberger2018numerical}), the focus has never been put on the full characterisation of the relaxation time of RBCs after the cessation of an external load, at changing systematically the strength of the mechanical load and the values of membrane viscosity. 
This paper aims at taking a step further in this direction.\\
So far, many numerical simulations have been done to investigate the mechanical properties of RBC membrane, and many models and methods have been developed: for the reader interested in these, we provide two reviews~\cite{fedosov2014multiscale,ju2015review}.
We exploit the combined Immersed Boundary - Lattice Boltzmann Method (IB-LBM) described in Kr\"{u}ger~\cite{thesis:kruger}, in which the elastic behaviour of RBCs is rendered through the Skalak model~\cite{art:skalaketal73} supplemented by a bending energy~\cite{book:gommperschick,thesis:kruger}. The quantitative details are accurately tuned by means of real RBCs properties. We account for membrane viscosity, as well, following the Standard Linear Solid model (SLS)~\cite{art:lizhang19}: the magnitude of membrane viscosity is kept as a variable parameter in the simulations. 
Our viscoelastic membrane model is quantitatively validated following a simple-to-complex trajectory: first, we study the deformation of a droplet in simple shear flow, whose elastic properties at the interface are given by surface tension; the latter is further supplemented by an interfacial viscosity. Then, in the same setup, we look at the deformation of a spherical capsule, whose interface is made by a thin membrane and its elastic behaviour is given by Skalak model; the latter is supplemented by membrane viscosity. Finally, we move to RBCs (note that, in our definition, a RBC is a capsule with a biconcave shape). The relaxation time of RBCs will then be quantitatively studied as a function of the membrane viscosity $\mum$ and the strength of mechanical loads imposed on the RBCs. We will show that implementing viscosity in the membrane model is crucial for capturing the dynamics of a real RBC. Moreover, we will perform simulations in two different setups: in the first one, we simulate the stretching of a single RBC in optical tweezers; in the second one, the RBC deforms under the action of an imposed shear flow. In general, it is found that the increase in the membrane viscosity results in a slower recovery dynamics. At fixed membrane viscosity, the systematic study at changing the strength and typology of mechanical load reveals other interesting properties: while in the limiting case of vanishing load strength the recovery dynamics is independent of the applied mechanical load, for finite load strengths it shows faster recovery dynamics with non universal features that markedly depend on the loading mechanism.\\
The paper is organised as follows. In Sec.~\ref{sec:method} we present the numerical model chosen for the RBC modelling. In Sec.~\ref{sec:benchmarks} some benchmark simulations against known literature data will be provided. Results on the relaxation properties of RBCs will be presented in Sec.~\ref{sec:rbc_results}. Conclusions will be drawn in Sec.~\ref{sec:conclusions}.
\section{Methodology: Immersed Boundary - Lattice Boltzmann Method (IB-LBM)} \label{sec:method}
\subsection{Numerical model for fluid solver and flow-structure coupling}\label{sec:fluid_solver}
\begin{figure*}
 \centering
 \includegraphics[width=.8\linewidth]{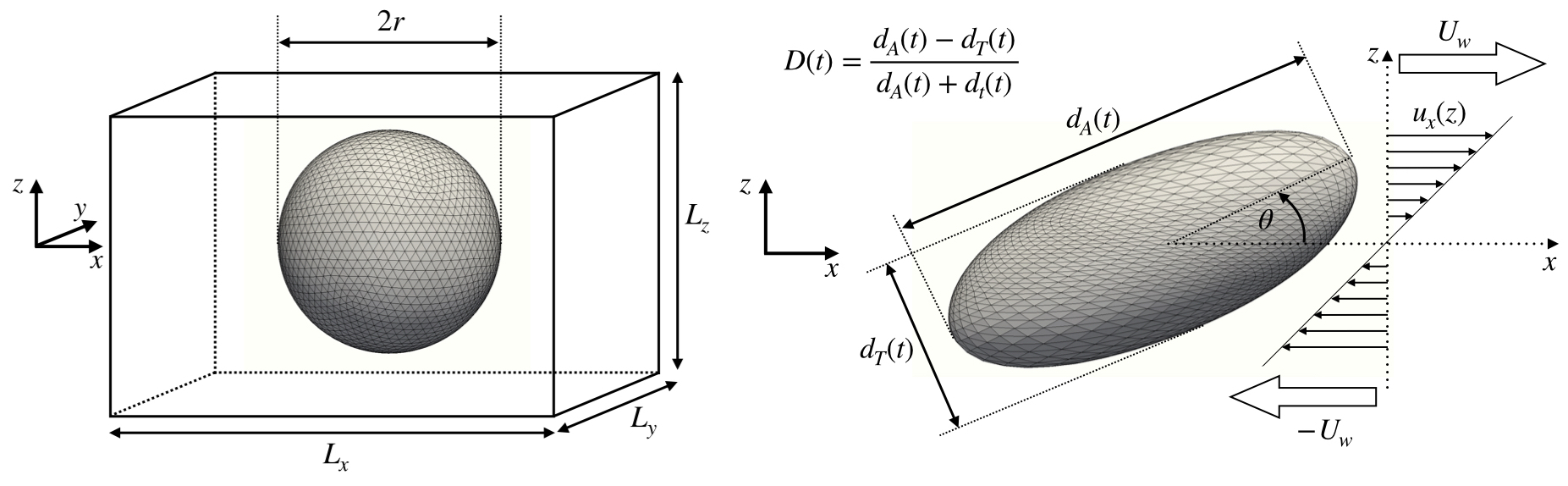}
\caption{Setup for the benchmarks on capsule/droplet deformation. The 3D mesh represents a spherical Lagrangian structure immersed in a fluid (cfr. Sec.~\ref{sec:fluid_solver}) in a computational box of size $\Lx \times \Ly \times \Lz$. The walls at $z=\pm \Lz/2$ move with opposite velocities $\pm \Uw$; $\Uw$ can be varied in order to impose a shear flow with prescribed intensity. Under the effect of the shear flow, the sphere deforms with major ($\dA$) and minor ($\dT$) axes in the shear plane: the deformation coefficient $D=(\dA-\dT)/(\dA+\dT)$ and the inclination angle $\theta$ are studied as functions of the flow properties. Different benchmark tests have been conducted: 
(i) Newtonian droplet deformation (cfr. Sec.~\ref{sec:bench_membrane_viscosity}) with additional surface viscosity; 
(ii) viscoelastic capsule deformation (cfr. Sec.~\ref{sec:viscoelastic_capsule}), for which the Lagrangian structure models both elastic and viscous properties of the capsule membrane.\label{fig:sphere}}
\end{figure*}

The hydrodynamical fields inside and outside RBCs are analysed through the Lattice Boltzmann Method (LBM), while the coupling with the Lagrangian structures is achieved via the immersed boundary (IB) method. The hybrid IB-LBM is already fully detailed in the literature~\cite{kruger2011efficient,book:kruger,thesis:kruger}: some essential features of the methodology are here recalled, while more extensive details can be found in dedicated books~\cite{book:kruger,succi2001lattice}.\\
LBM is a numerical approach which is grounded on the discretisation of Boltzmann kinetic equation~\cite{book:kruger,succi2001lattice}, rather than on the continuum assumption, which is at the basis of Navier-Stokes approach. LBM hinges on the dynamical evolution of fluid molecules distribution functions, $\effi(\vec{x},t)$, which represent the density of fluid molecules with velocity $\vec{c}_i$ at position $\vec{x}$ and time $t$:
\begin{multline}\label{LBMEQ}
\effi(\vec{x}+\vec{c}_i\Delta t, t+ \Delta t) - \effi(\vec{x}, t) =\\
=-\frac{\Delta t}{\tau}\left(\effi(\vec{x}, t) - \effi^{(\mbox{\tiny eq})}(\vec{x}, t)\right) + \effi^{(F)}
\end{multline}

in which $\tau$ is the relaxation time, $\effi^{(F)}$ is the force density (which has been implemented according to the so-called ``Guo'' scheme~\cite{PhysRevE.65.046308}) and $\effi^{(\mbox{\tiny eq})}(\vec{x}, t)$ is the equilibrium distribution function~\cite{succi2018lattice}:
\begin{equation}\label{eq_pop}
\effi^{(\mbox{\tiny eq})} = w_i\rho\left(1+\frac{\vec{c}_{i}\cdot \vec{u}}{\cs^2} + \frac{\left(\vec{c}_{i}\cdot \vec{u}\right)^2}{\cs^4} - \frac{\vec{u}\cdot \vec{u}}{\cs^2} \right) \; .
\end{equation}
In Eq.~\eqref{eq_pop}, $\rho$ is the fluid density and $\cs$ is the speed of sound. In the velocity set $\{c_i\}$ that we have chosen, it results $\cs^2 =  (1/3)(\Delta x/ \Delta t)^2$.
Note that, since we are interested in simulating two different fluids (one outside and one inside the membrane of a capsule), the relaxation time is not constant but it depends on the position, i.e. $\tau = \tau(\vec{x})$.\\
From the evolution of the distribution function $\effi(\vec{x}, t)$, we retrieve the most important (macroscopic) hydrodynamic quantities, that are the density $\rho$ and the velocity of the fluid $\vec{u}$:
\begin{equation}
\rho(\vec{x}, t) = \sum_{i} \effi(\vec{x}, t)\; , \qquad \rho\vec{u}(\vec{x}, t) = \sum_{i} \vec{c}_i \effi(\vec{x}, t) 
\end{equation}
where the sum runs over the discrete directions. 
\\
The link between the Navier-Stokes equation and the LBM, given by the Chapman-Enskog analysis, provides the relation between the LB relaxation time $\tau$ and the viscosity of the fluid $\muout$~\cite{book:kruger}:
\begin{equation}
\muout = \rho \cs^2\left(\tau-\frac{\Delta t}{2}\right)\; .
\end{equation}
We also define the viscosity ratio $\viscratio$ as the ratio between the inner fluid and outer fluid viscosities:
\begin{equation}
\viscratio = \frac{\muin}{\muout}\; .
\end{equation}
In this work, we use LBM to simulate flows in the Stokes limit (see Sec.~\ref{sec:bench_elastic}), that is for low Reynolds number $\Rey=\dot{\gamma}\rho r^2/ \muout$ , where $\dot{\gamma}$ is the shear rate and $r$ is the characteristic length, i.e. the radius of the spherical capsule/droplet (see Fig.~\ref{fig:sphere}) or the radius of the RBC at rest (see Fig.~\ref{fig:rbc}).\\
The fluid dynamics of LBM is further coupled with Lagrangian structures (such as the interface of a droplet or the membrane of a capsule) via the IB method~\cite{book:kruger}. The Lagrangian structure is represented by a finite number of Lagrangian points that can freely move through the Eulerian lattice. The velocity $\dot{\vec{r}}_j$ of the $j$-th Lagrangian point at position $\vec{r}_j$ is computed by interpolating the fluid velocity $\vec{u}$ of the neighbouring lattice sites $\vec{x}$ by means of discrete Dirac delta functions $\Delta(\vec{r}_j(t),x)$:
\begin{equation}
\dot{\vec{r}}_j(t) = \sum_{\vec{x}}\Delta x^3\vec{u}(\vec{x},t)\Delta(\vec{r}_j(t),\vec{x})\; .
\end{equation}
The nodal force $\vec{F}_j$ on the $j$-th Lagrangian node is computed and \textit{passed} to the neighbouring lattice sites with the same interpolation adopted to compute the nodal velocity. 
\begin{equation}
\vec{F}(\vec{x},t) = \sum_j\vec{F}_j(t)\Delta(\vec{r}_j(t),\vec{x})\; .
\end{equation}
Such a nodal force $\vec{F}_j$ is strongly related to the adopted membrane model (see Sec.~\ref{sec:membrane_model}). The Dirac delta function can be factorised in the interpolation stencils, $\Delta(\vec{x})=\phi(x)\phi(y)\phi(z)/\Delta x^3$ \cite{book:kruger,kruger2011efficient}. Unless differently specified, we adopt the 4-point interpolation stencil:
\begin{equation}
\small\phi_4(x) = \begin{cases}
\frac{1}{8}\left(3-2\vert x\vert + \sqrt{1+4\vert x\vert-4x^2} \right) & 0\le \vert x\vert\; ;\\
\frac{1}{8}\left(5-2\vert x\vert - \sqrt{-7+12\vert x\vert-4x^2} \right) & \Delta x\le \vert x\vert\le 2\Delta x\; ; \\
0 & 2\Delta x\le \vert x\vert\; .
\end{cases}
\end{equation}
Finally, in order to recognise which lattice sites are enclosed by the Lagrangian structure at each time step, we have implemented the parallel Hoshen-Kopelman algorithm~\cite{art:frijters15}. Briefly, each processor first recognises the lattice sites near the Lagrangian structure, and by using the outgoing normal for each face it distinguishes which of these sites are inner and which outer; then, it assigns different labels for lattice sites in different clusters; finally, all the processors communicate the labels and by looping over the lattice domain they recognise whether a lattice site is inside or outside the Lagrangian structure. Extensive details can be found in~\cite{art:frijters15}.  

%
\subsection{Membrane model}\label{sec:membrane_model}
The configurational properties of the membrane are defined through its free energy $W$, whose main contributions are the bending~\cite{book:gommperschick,art:helfrich73} and the  strain~\cite{art:skalaketal73} energies. These two terms are implemented as proposed in~\cite{thesis:kruger}. We further introduce the membrane viscosity, by employing the algorithm proposed by Li and Zhang \cite{art:lizhang19}. In the following Subsections, we recall the main ingredients of the aforementioned implementations.
\subsection*{Bending and Strain}\label{sec:bendingElasticity}
The bending term is introduced to take into account that the RBC has a bilayer membrane. We adopt Helfrich formulation to describe the bending resistance~\cite{book:gommperschick,art:helfrich73}. The free energy $\WB$ is given by~\cite{thesis:kruger}:
\begin{equation}\label{eq:helfrich1}
\WB = \frac{\kB}{2}\int_\Sigma \left(H-H^{(0)}\right)^2 d\Sigma
\end{equation}
where $H$ is the trace of the surface curvature tensor, $H^{(0)}$ is the spontaneous curvature and $\Sigma$ is the capsule surface area; $\kB$ is the {\it bending modulus}, whose value is reported in the table of Fig.~\ref{fig:rbc}~\cite{book:gommperschick}. Following~\cite{thesis:kruger}, we discretise Eq.~\eqref{eq:helfrich1} as follows:
\begin{equation}\label{eq:helfrich2}
\WB = \frac{\kB\sqrt{3}}{2}\sum_{\langle i,j\rangle}\left(\theta_{ij}-\theta_{ij}^{(0)}\right)^2
\end{equation}
where the sum runs over all the neighbouring triangular elements, and $\theta_{ij}$ is the angle between the normals of the $i$-th and $j$-th elements (note that $\theta_{ij}^{(0)}$ is the same angle in the unperturbed configuration).\\
The strain contribution to the total free energy can be written as~\cite{art:skalaketal73}
\begin{equation}\label{eq:skalak1}
\WS = \sum_j w_S^{(j)}A_j
\end{equation}
where $w^{(j)}_S$ is the area energy density related to the $j$-th element with surface area $A_j$~\cite{art:skalaketal73}
\begin{equation}\label{eq:skalak2}
w^{(j)}_S= \frac{\kS}{8}\left(I_1^2+2I_1-2I_2\right) +  \frac{\Kal}{8} I_2^2
\end{equation}
where $I_1 = \lambda_1^2+\lambda_2^2-2$ and $I_2 = \lambda_1^2\lambda_2^2-1$ are the strain invariants for the $j$-th element, while $ \lambda_1$ and $ \lambda_2$ are the principal stretch ratios~\cite{art:skalaketal73,thesis:kruger}: note that $\lambda_1$ and $\lambda_2$  are the eigenvalues of the deformation gradient tensor $\vec{\mathcal{F}}$. The coefficients $\kS$ and $\Kal$ are the {\it surface elastic shear modulus} and the {\it area dilation modulus}, respectively: their physical values for the RBC membrane are reported in the table of Fig.~\ref{fig:rbc}~\cite{art:suresh2005connections,book:gommperschick}. Moreover, $\kS$ is related to the Capillary number \Ca, defining the relative importance of bulk viscous effects with respect to the elastic properties of the membrane:
\begin{equation}\label{eq:ca_capsule}
\Ca = \frac{\muout\dot{\gamma}r}{\kS}\; .
\end{equation}
Once the total free energy $W = \WB + \WS$ is computed, the force acting on the $i$-th node at the position $\vec{x}_i$ is retrieved:
\begin{equation}\label{eq:nodal_force_energy}
\vec{F}_i = -\frac{\partial W(\vec{x}_i)}{\partial \vec{x}_i}\; .
\end{equation}
\subsection*{Membrane viscosity}\label{sec:membrane_viscosity}
The viscosity of the membrane relates the 2D viscous stress $\vec{\tau}^\nu$ to the strain rate tensor $\dot{\vec{E}}$, i.e. the time derivative of the strain tensor:
\begin{equation}\label{eq:strain_tensor}
\vec{E} = \frac{1}{2}\left( \vec{F}^T\vec{F} - \mathbb{1}\right)
\end{equation}
where $ \mathbb{1}$ is the 2-dimensional unit matrix. Two contributions can be identified in the strain rate tensor $\dot{\vec{E}}$, related to the shear $(\bullet_s)$ and the dilatational  $(\bullet_d)$ contributions, respectively:
\begin{equation}
\dot{\vec{E}}_s = \dot{\vec{E}} - \frac{1}{2}\mbox{tr}(\dot{\vec{E}})\mathbb{1}\; ,\hspace{.5in} \dot{\vec{E}}_d = \frac{1}{2}\mbox{tr}(\dot{\vec{E}})\mathbb{1}\; .
\end{equation}
By introducing the shear and dilatational membrane viscosities ($\mus$ and $\mud$, respectively), we can split the 2D viscous stress $\vec{\tau}^\nu$ in two parts:
\begin{multline}\label{eq:mv2}
\vec{\tau}^\nu = \vec{\tau}_s^\nu +\vec{\tau}_d^\nu = 2\mus \dot{\vec{E}}_s + 2\mud\dot{\vec{E}}_d =\\= \mus \left(2\dot{\vec{E}} -\mbox{tr}(\dot{\vec{E}})\mathbb{1}\right) + \mud \mbox{tr}(\dot{\vec{E}})\mathbb{1}\; .
\end{multline}
We employ the standard linear solid (SLS) model to describe the viscoelastic behaviour of the membrane~\cite{art:lizhang19,art:yazdanibagchi13}: it can be considered as the linear combination of the Kelvin-Voigt model (in which a dashpot $\mu$ is connected in parallel with a spring $k$) and the Maxwell model (in which the dashpot $\mu$ and an artificial spring $k'$ are connected in series). The Kelvin-Voigt model can successfully describe membrane creep, while the Maxwell model accurately characterises the stress relaxation; their combination, the SLS model, correctly accounts for both phenomena~\cite{art:lopezguerra14}. The value of $k'$ has been tuned in order to retrieve the asymptotic behaviour of the Maxwell element~\cite{art:lizhang19}. We refer the interested reader to previous literature papers~\cite{art:lopezguerra14,art:yazdanibagchi13,art:lizhang19} for extensive technical details on the model. The membrane viscosity properties are quantified via the Boussinesq numbers: 
\begin{equation}\label{eq:bq_capsule}
\Bqs=\frac{\mus}{\muout r}\; , \qquad\qquad \Bqd=\frac{\mud}{\muout r} 
\end{equation}
accounting for shear and membrane dilation effects, respectively. Once the total stress $\vec{\tau}^T$  has been computed, the total force acting on the $i$-th node is given by  
\begin{equation}\label{eq:nodal_force}
\left(\begin{matrix}
F_x^i \\ F_y^i
\end{matrix} \right)=\vec{\tau}^T \vec{\mathcal{F}}^{-T}\nabla N_i A_m = 
(\vec{\tau}^\nu+\vec{\tau}^s) \vec{\mathcal{F}}^{-T} \left(\begin{matrix}
a_i \\ b_i
\end{matrix} \right) A_m
\end{equation}
where $N_i = a_ix+b_iy+c_i$ is the linear shape function of the $m$-th element, and $A_m$ is the surface area of the $m$-th element (see~\cite{art:shrivastava93} for more details). 
For the droplet with interfacial viscosity (Sec.~\ref{sec:bench_membrane_viscosity}), the total stress $\vec{\tau}^T$ is the sum of the viscous stress $\vec{\tau}^\nu$ and the surface tension stress  $\vec{\tau}^s$; while for the viscoelastic capsule (Sec~\ref{sec:viscoelastic_capsule}), and then also for the RBC (Sec.~\ref{sec:rbc_results}), since the elastic force is directly computed from the free energy (cfr. Eq.~\eqref{eq:nodal_force_energy}), the total stress is $\vec{\tau}^T = \vec{\tau}^\nu$.
%

\begin{figure*}[th!]
\centering
\centering\includegraphics[width=.4\linewidth]{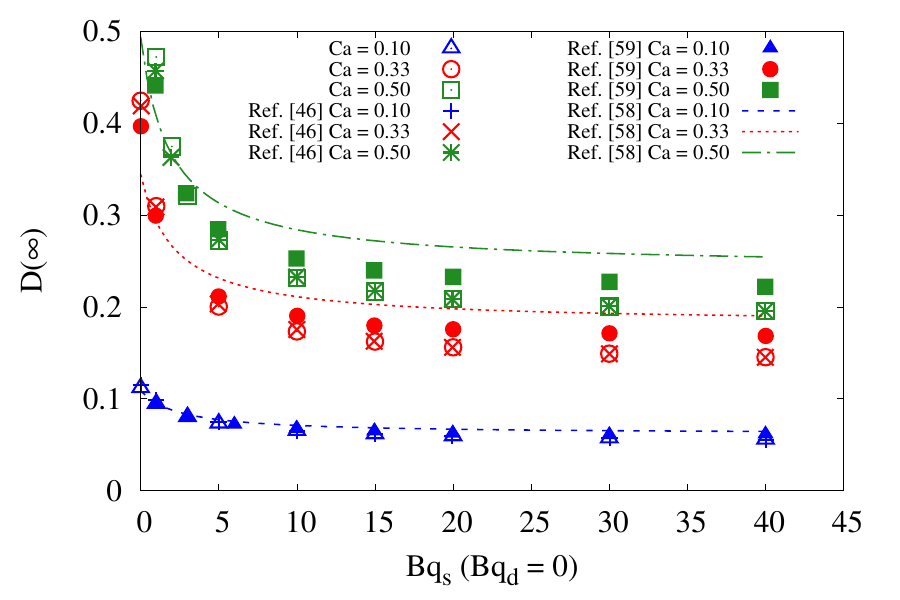}
\centering\includegraphics[width=.4\linewidth]{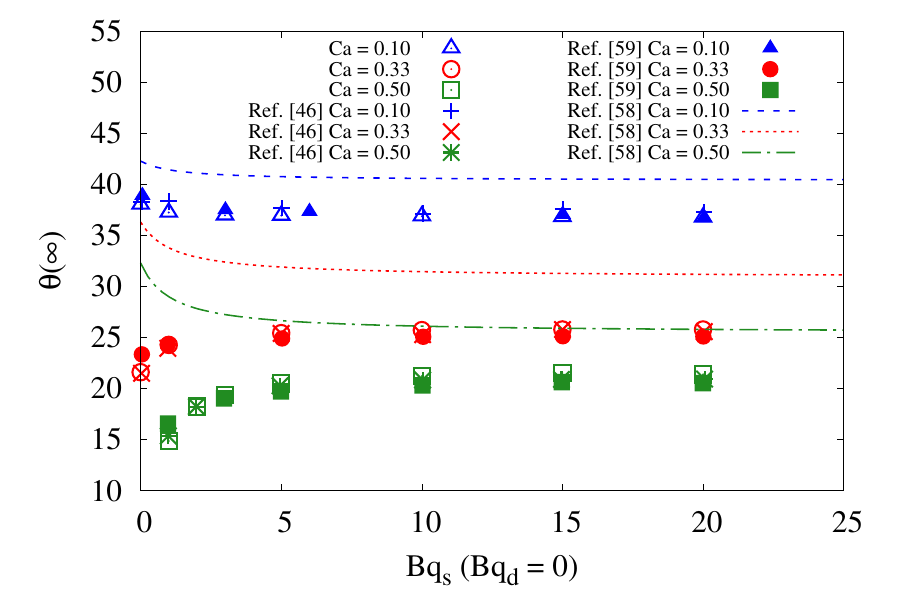}
\centering\includegraphics[width=.4\linewidth]{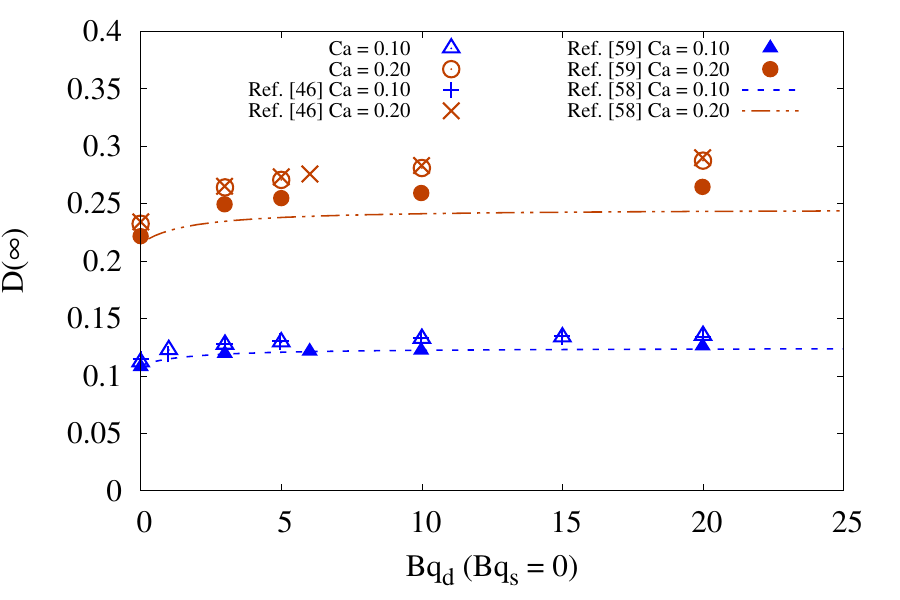}
\centering\includegraphics[width=.4\linewidth]{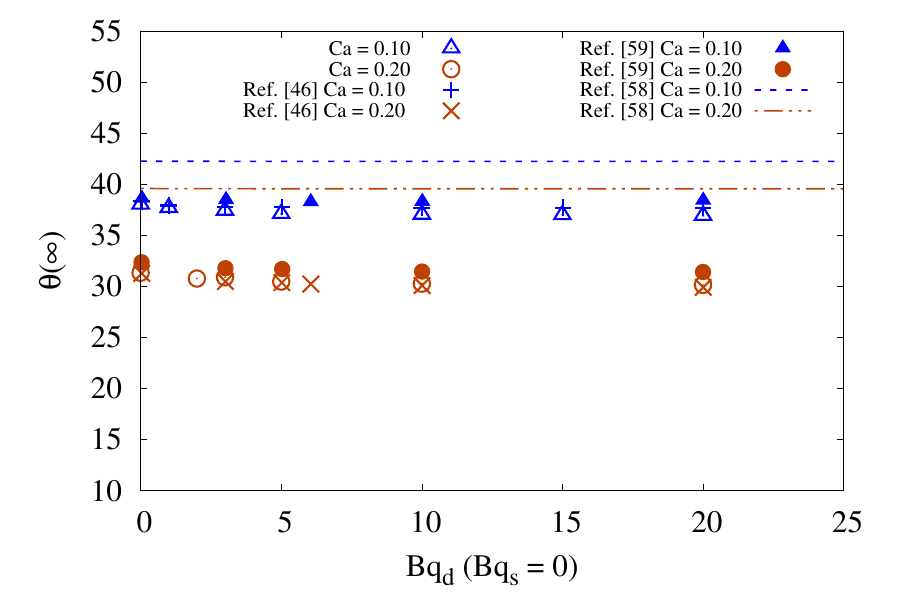}
\centering\includegraphics[width=.4\linewidth]{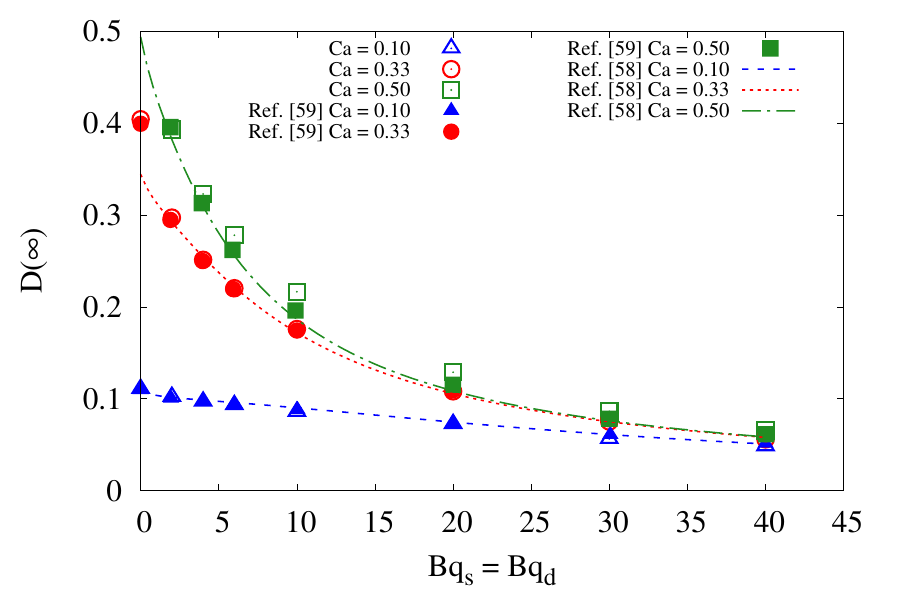}
\centering\includegraphics[width=.4\linewidth]{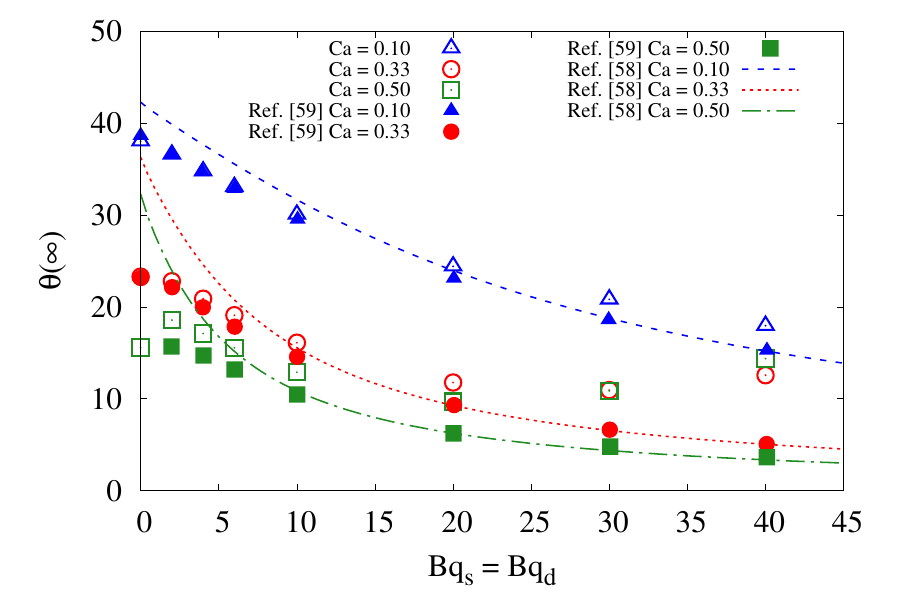}
\caption{Deformation of Newtonian droplet with interfacial viscosity in simple shear flow (see Fig.~\ref{fig:sphere}). We report the steady deformation $D(\infty)$ (Left panels) and inclination angle $\theta(\infty)$ (Right panels) as functions of the Capillary number $\Ca= \frac{\muout \dot{\gamma}r}{\varsigma}$, for different values of the Boussinesq numbers $\Bqs$ and   $\Bqd$ (cfr. Eq.~\eqref{eq:bq_capsule}). The results from numerical simulations are compared with the perturbative theoretical calculations by Flumerfelt~\cite{art:flumerfelt80}, the Boundary Integral calculations by Gounley {\it et al. }~\cite{art:gounley16} and the IB-LBM numerical simulations by Li \& Zhang~\cite{art:lizhang19}.}
\label{fig:bench_droplet}
\end{figure*}
\section{Benchmarks}\label{sec:benchmarks}
To benchmark the numerical model, we conducted 3D simulations of an initially spherical Lagrangian structure (droplet or capsule) deforming in an imposed linear shear flow, within a domain of size $\Lx \times \Ly \times \Lz$. Through the bounce-back method for moving walls~\cite{book:kruger}, we set the wall velocity $\Uw=(\pm \dot{\gamma} \Lz/2,0,0)$ on the two plane walls at $z=\pm \Lz/2$ ($\Uw>0$ for $z>0$); periodic boundary conditions are set along $x$ and $y$ directions. The spherical structure will deform under shear, developing an inclination angle $\theta$ with respect to the flow direction (see Fig.~\ref{fig:sphere}). We define the deformation parameter $D$ as 
\begin{equation}\label{eq:DEFORMATION}
D = \frac{\dA-\dT}{\dA+\dT}
\end{equation}
where $\dA$ and $\dT$ are the major and minor axes of the structure in the shear plane (see Fig.~\ref{fig:sphere}). We determine $\dA$ and $\dT$ via the three real eigenvalues $T_1, T_2$ and $T_3$ of the inertia tensor $\vec{T}$ as $\dA = 2\sqrt{5\left(T_2+T_3-T_1\right)/(2M)}$ and $\dT = 2\sqrt{5\left(T_3+T_1-T_2\right)/(2M)}$, where $M$ is the mass of the immersed object~\cite{thesis:kruger}. To compute the inclination angle $\theta$, we consider the eigenvectors of $\vec{T}$.\\ 
To test both the elastic and the viscous models, we follow a simple-to-complex path: first, we consider a droplet, whose elastic response of the interface is given by the surface tension further supplemented by an interfacial viscosity implemented with the SLS model (see Sec.~\ref{sec:membrane_model}). Then, we consider a spherical capsule, whose interface is made by a thin membrane and its elastic properties are given by the Skalak model (see Sec.~\ref{sec:membrane_model}): for the spherical capsule, we first test only the Skalak model (i.e. without membrane viscosity); then, we implement the SLS model that has been already tested for the droplet. At the end of this section, we have a fully validated viscoelastic model that we use in Sec.~\ref{sec:rbc_results} to simulate e realistic RBC.\\
In these benchmarks, we represent both the droplet and the spherical capsule (see Fig.~\ref{fig:sphere}) via a 3D triangular mesh generated by a successive subdivision of a regular icosahedron.\\ 

\begin{figure}[t!]
\centering
\includegraphics[width=.9\linewidth]{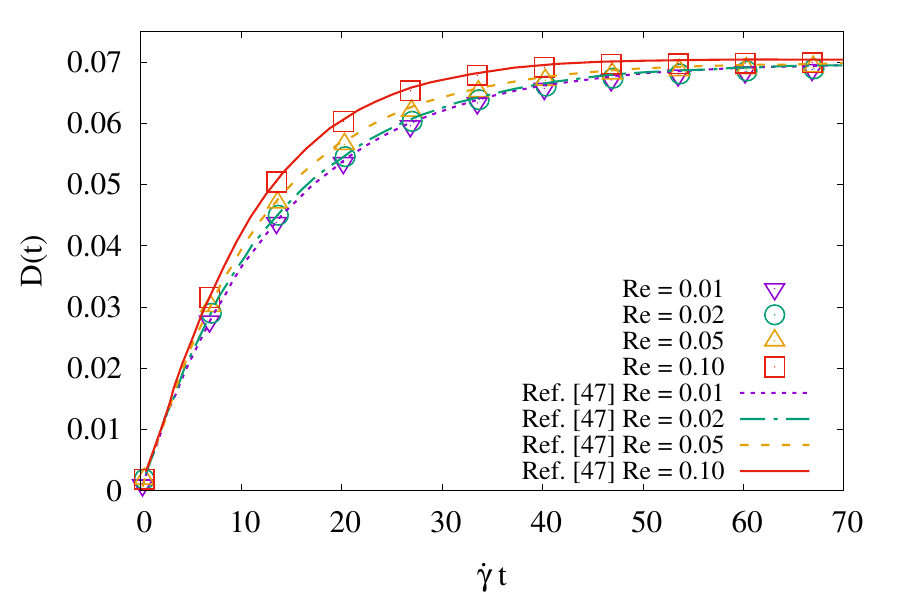}
\includegraphics[width=.9\linewidth]{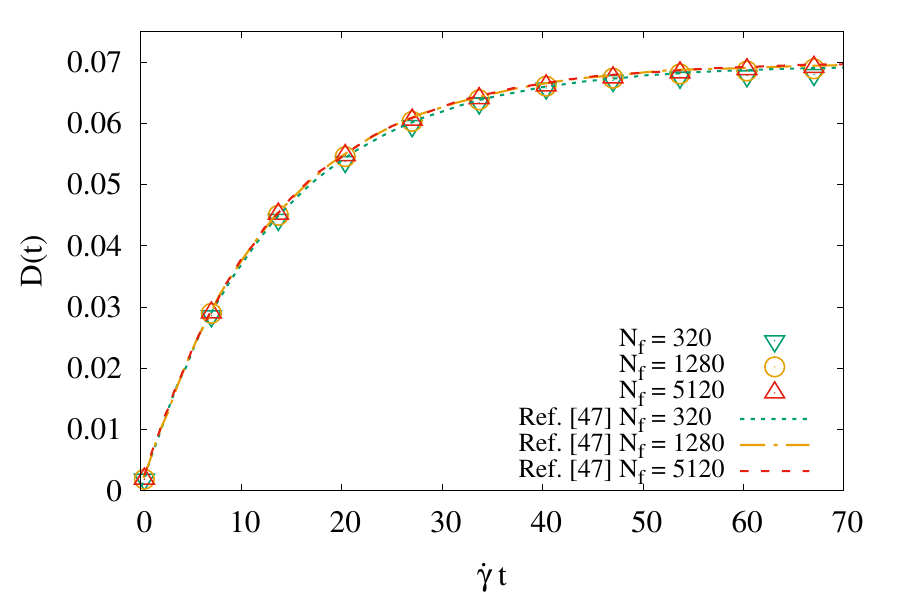}
\caption{Deformation of purely elastic capsule in simple shear flow (see Fig.~\ref{fig:sphere}). Top panel: we report the deformation parameter $D(t)$ (cfr. Eq.~\eqref{eq:DEFORMATION}) as a function of the dimensionless time $\dot{\gamma}t $, with $\dot{\gamma}$ the imposed shear rate; the Reynolds number $\Rey$ is varied to test the convergence to the Stokes' limit. Bottom panel: we report $D(t)$ as a function of the dimensionless time $\dot{\gamma}t$ for a fixed Reynolds number $\Rey=0.01$. In the panels, the trends are functions of the the number of faces $\Nf$ used to discretise the 3D surface of the elastic sphere. Comparisons with data by Kr\"{u}ger {\it et al.}~\cite{kruger2011efficient} (lines) are performed. \label{fig:kruger_bench}}
\end{figure}

\subsection{Newtonian droplet with interfacial viscosity}\label{sec:bench_membrane_viscosity}
In order to benchmark the implementation of membrane viscosity, we have investigated the deformation of a Newtonian droplet in shear flow, including the contribution of the surface viscosity at the droplet interface. The total stress on the interface is given by the sum of the viscous tensor $\vec{\tau}^\nu$ \eqref{eq:mv2} and is supplemented with the surface tension effects: these can be introduced by adding to the viscous stress tensor the surface tension tensor $\vec{\tau}^s~=~\varsigma \mathbb{1}$~\cite{art:lizhang19}, with $\varsigma$ the surface tension coefficient. Such a form of the surface tension tensor is equivalent to a force per unit area equal to minus the derivative of the surface free energy $w_{\varsigma} = \varsigma \int_\Sigma d\Sigma$. In order to quantify the response under a simple shear flow, we studied the steady state deformation of the droplet $D(\infty)$ and the inclination angle $\theta(\infty)$ (see Fig.~\ref{fig:sphere}) by varying the Capillary number $\Ca = \frac{\muout \dot{\gamma}r}{\varsigma}$, and the Boussinesq numbers $\Bqs$ and $\Bqd$ (cfr. Eqs.~\eqref{eq:bq_capsule}), following the benchmarks proposed by Li \& Zhang~\cite{art:lizhang19}. Thus, we adopt the same numerical parameters as in~\cite{art:lizhang19} and we further extend the validation by considering various choices of the Boussinesq numbers. We will also provide comparisons with the boundary integral calculations by Gounley {\it et al.}~\cite{art:gounley16} and the theoretical predictions by Flumerfelt~\cite{art:flumerfelt80}. The results of our simulations are reported in Fig.~\ref{fig:bench_droplet}. For fixed choice of $\Bqs$ and $\Bqd$ the deformation increases with the Capillary number, while the orientation angle decreases. At fixed Capillary number, the choices $\Bqs=\Bqd$ and $\Bqs \neq 0$ and $\Bqd = 0$ correspond to a decreasing deformation at increasing Boussinesq number; at contrary, the case $\Bqs=0$ and $\Bqd \neq 0$ corresponds to an increasing deformation, which may seem as counterintuitive. However, one has to note that in such a case the overall viscous stress is proportional to the divergence of the velocity field: the numerical simulations by Gounley {\it et al.}~\cite{art:gounley16} actually show that the surface velocity divergence becomes negative in the fluid regions at the ends of the major axes and this is instrumental to lower the stress in such regions and promote larger droplet deformations.\\ 
Overall, the results of our simulations show an excellent agreement with those of Li \& Zhang~\cite{art:lizhang19}. Some discrepancies emerge in the comparisons with the results of Flumerfelt~\cite{art:flumerfelt80} and Gounley {\it et al.}~\cite{art:gounley16}. One has to notice, however, that Flumerfelt's calculations are perturbative in $\Ca$, hence they are not suited to match the regimes at larger $\Ca$, by definition. Moreover, while our calculations and those by Li \& Zhang~\cite{art:lizhang19} are based on the LBM, Gounley {\it et al.}~\cite{art:gounley16} solve the Stokes equations by means of the boundary integral method. Discrepancies between the LBM and boundary integral formulation were already observed in Li \& Zhang~\cite{art:lizhang19}; here, thanks to the systematic benchmarks at changing Boussinesq numbers, we can see that such discrepancies are not observed when studying the steady state deformation with the condition $\Bqs=\Bqd$: with such a choice, the term $\mbox{Tr}(\dot{\vec{E}})$ disappears from the 2D viscous stress $\vec{\tau}^\nu$ (cfr. Eq.~\eqref{eq:mv2}). Still, when $\Bqs=\Bqd$, the orientation angles differ from those presented by Gounley {\it et al.}~\cite{art:gounley16} of a few degrees. Such difference may originate from the way we compute the inclination angle, i.e. we get the information on the inclination angle from the eigenvectors of $T$. In fact, if we consider the radius which connect the center of the triangular element to the center of mass, a small variation in the coordinates of this element can result in a variation of the inclination of the radius of a few \%. Hence, increasing the number of elements used to describe the mesh may improve the accuracy. The above considerations point to the fact that the observed dissimilarities with Gounley {\it et al.}~\cite{art:gounley16} can originate from surface divergence terms of the velocity field. Further analysis is obviously needed to better address and quantify such discrepancies. In the following, RBC simulations will be performed with the choice $\Bqs=\Bqd$~\cite{barthes1985}.
%
\begin{figure}[t!]
\centering
\includegraphics[width=.9\linewidth]{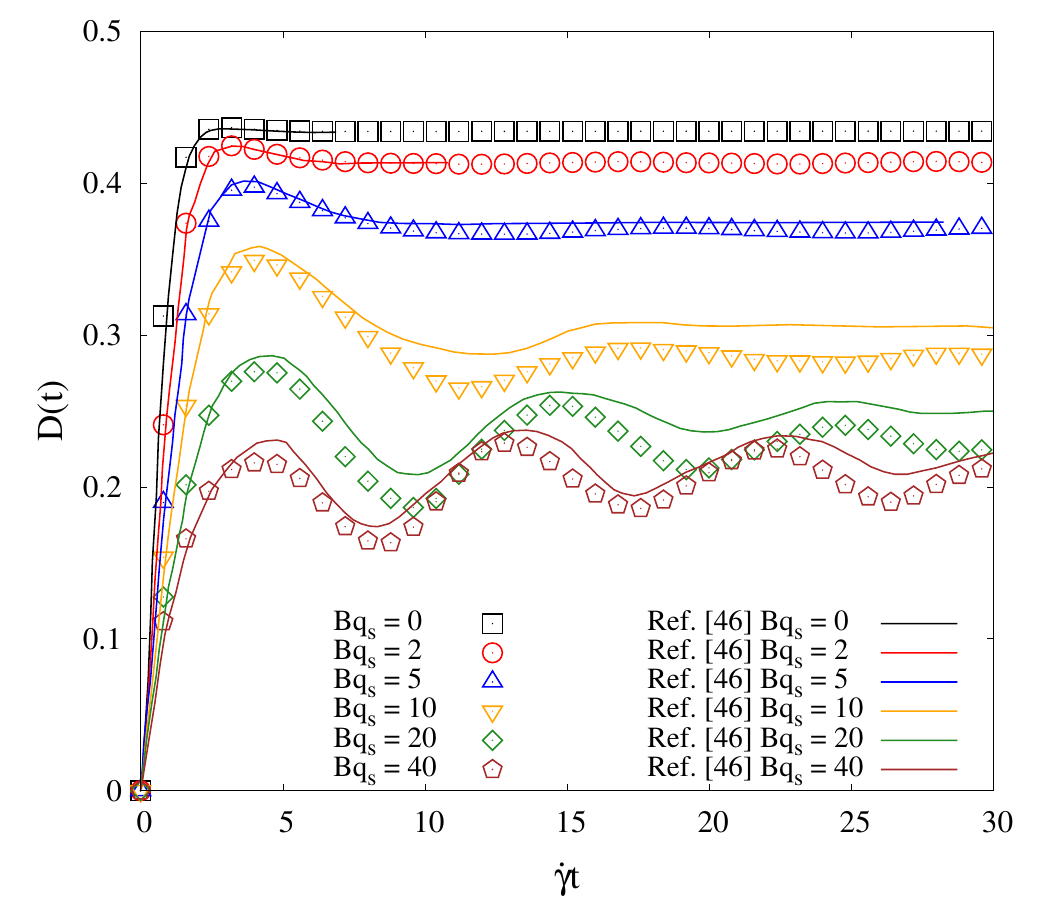}
\caption{Deformation of a viscoelastic capsule in simple shear flow with prescribed intensity of the shear rate $\dot{\gamma}$ (see Fig.~\ref{fig:sphere}). We report the time evolution of the deformation parameter $D(t)$ (cfr. Eq.~\eqref{eq:DEFORMATION}) as a function of the dimensionless time $\dot{\gamma}t$. We vary the Boussinesq number $\Bqs$ while keeping $\Bqd=0$ (cfr. Eqs.~\eqref{eq:bq_capsule}). Results from numerical simulations are compared with the IB-LBM data provided by Li \& Zhang~\cite{art:lizhang19}.\label{fig:li_fig7}}
\end{figure}

\begin{figure*}[t!]
\centering
\begin{minipage}{0.48\linewidth}
\includegraphics[width=.8\linewidth]{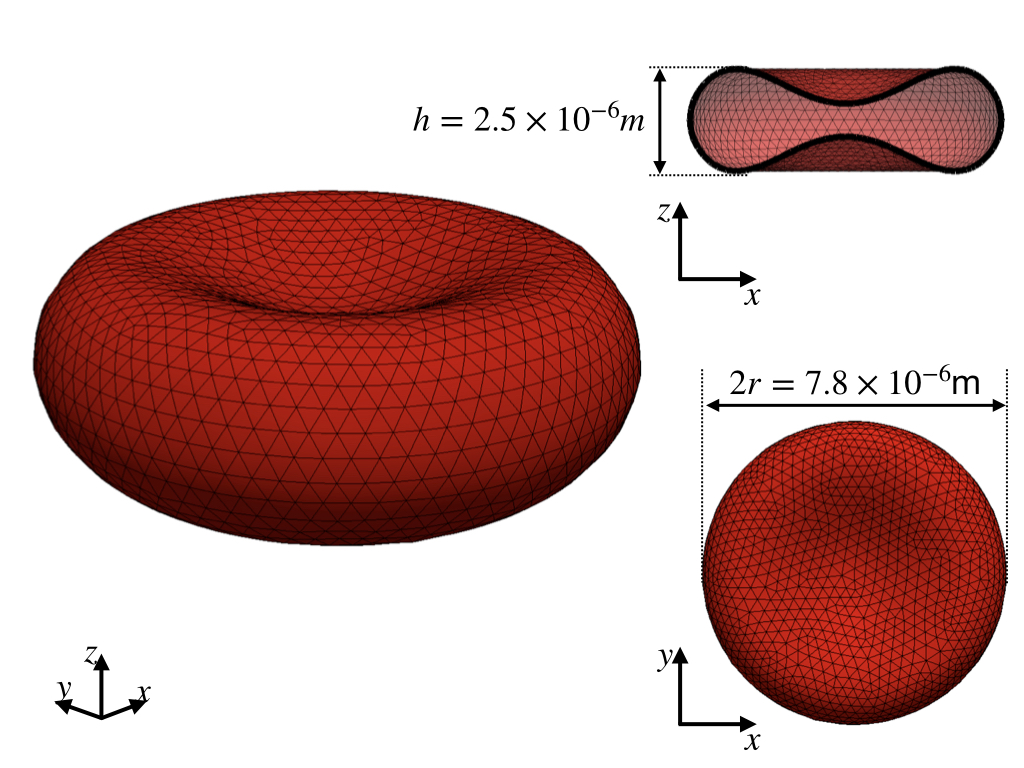}
\end{minipage}
\begin{minipage}{0.48\linewidth}
\begin{tabular}{ll}
\hline
\multicolumn{2}{c}{RBC PARAMETERS} \\
\hline
Radius $r$ & $3.91\times 10^{-6}$ m\ \ ~\cite{thesis:kruger} \\
Area $A$ & $133\times 10^{-12}\mbox{ m}^2$ \\
Volume $V$ & $93\times 10^{-18}\mbox{ m}^3$ \\
Elastic shear modulus $\kS$ & $5.3\times10^{-6}\mbox{ Nm}^{-1}$\ \ ~\cite{art:suresh2005connections} \\
Elastic dilatational modulus $\Kal$ & 50 $k_s$\ \ ~\cite{thesis:kruger}  \\
Bending modulus $\kB$ & $2\times 10^{-19}\mbox{ Nm}$\ \ ~\cite{book:gommperschick} \\
Plasma viscosity $\muout$ & $1.2\times 10^{-3}$ Pa s\ \ ~\cite{thesis:kruger}\\
Cytoplasm viscosity $\muin$ & $6\times 10^{-3}$ Pa s\ \ ~\cite{thesis:kruger} \\
\hline
\end{tabular}
\end{minipage}
\caption{3D triangular mesh used to represent the viscoelastic membrane of the RBC: the elastic behaviour is described by the Skalak model, while the Standard Linear Solid Model (SLS) is used to take into account the membrane viscosity (see Sec.~\ref{sec:membrane_model}). The table on the right summarises the physical dimensions, the elastic parameters and the adapted transport coefficients.\label{fig:rbc}}
\end{figure*}

\subsection{Capsule with viscoelastic membrane}\label{sec:viscoelastic_capsule}\label{sec:bench_elastic}
First, to benchmark the Skalak model (see Sec.~\ref{sec:membrane_model}), we simulated the deformation of a purely elastic capsule (i.e. with no membrane viscosity) immersed in a linear shear flow: therefore, we use only the free-energy term related to the strain and area dilation given by Eq.~\eqref{eq:skalak2}. To benchmark our implementation, we performed the same set of simulations as those proposed by Kr\"{u}ger {\it et al.}~\cite{kruger2011efficient}, by varying the Reynolds number $\Rey$, the number of faces $\Nf$, the aspect ratio $\Lz/r$ and the Capillary number $\Ca$: all the adopted parameters have exactly the same values as those in~\cite{kruger2011efficient}. Some representative results are displayed in Fig.~\ref{fig:kruger_bench}. In the top panel, we show the time evolution of the deformation parameter at changing the Reynolds number $\Rey$. The plot shows a neat asymptotic limit when $\Rey$ decreases (Stokes' limit). The results indicate that $\Rey=0.01$ is small enough to achieve a fair convergence. In the bottom panel, we show the time evolution of the deformation parameter $D(t)$ as a function of the number of faces $\Nf$ used to discretise the spherical capsule. A proper choice of $\Nf$ is, in fact, crucial both to avoid numerical errors due to a coarse mesh and to ensure the Hoshen-Kopelman algorithm to properly work~\cite{art:frijters15}. Results indicate that $\Nf=5120$ is sufficient to make the numerical errors related to the mesh resolution negligible. 
As a further benchmark (results shown in the Electronic Supplementary Information (ESI) section), we have checked the deformation and the inclination angle by varying the LB relaxation time $\tau$, the radius of the sphere $r$, the aspect ratio $\Lz/r$ and the interpolation stencil $\phi$, finding an excellent agreement with the results presented in Kr\"{u}ger {\it et al.}~\cite{kruger2011efficient}, thus supporting the accuracy and rigor of the implemented methodology.\\
Finally, we benchmark the numerical simulations in the presence of both elasticity and membrane viscosity. The reference data are those presented in Li \& Zhang~\cite{art:lizhang19}. We adopted the very same numerical parameters as those in~\cite{art:lizhang19}; in particular, the elastic properties of the membrane are described by the Skalak model, with $\Kal=\kS$ (cfr. Eq.~\eqref{eq:skalak2}), supplemented with membrane viscosity with variable $\Bqs$ and $\Bqd=0$. In Fig.~\ref{fig:li_fig7}, we report the deformation $D(t)$ as a function of the dimensionless time $\dot{\gamma}t$: this figure highlights the satisfactory agreement with the results by Li \& Zhang~\cite{art:lizhang19}. In particular, oscillations in the transient dynamics are observed, which are dumped at large simulation times. The persistence of the oscillations is more pronounced with increasing $\Bqs$, because the response time of the capsule becomes much larger that the flow time scale $\dot{\gamma}^{-1}$~\cite{art:lizhang19,art:yazdanibagchi13}. Notice that due to the absence of the bending term, some wrinkles may appear on the surface~\cite{art:lizhang19}, and the shape of the capsule can be slightly different from an ellipsoid. This could be the cause of the very slight misalignment between the two sets of data. 
%

\begin{figure*}[th!]
\centering
\includegraphics[width=.8\linewidth]{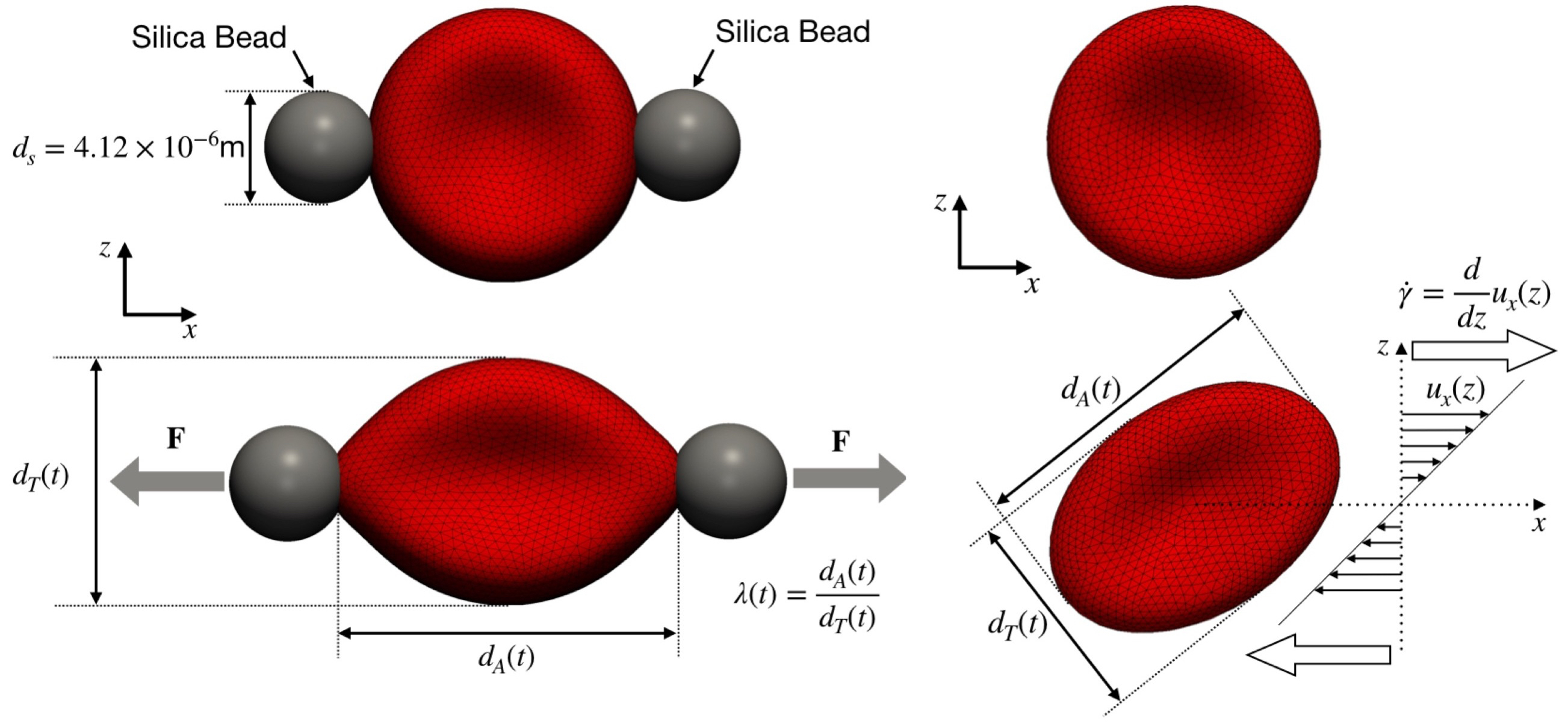}
\caption{Deformation of a single RBC under an external mechanical load. Left panel: stretching caused by 2 forces with the same intensity $F$ and opposite direction, applied at the ends of the erythrocyte. Forces are applied on a selected number of nodes (see text for details), in order to mimic the experimental stretching of the RBC induced by two silica beads bonded to the cell ends in an optical tweezer~\cite{art:suresh2005connections}. The RBC deformation is quantified with the extension ratio $\lambda(t)=\dA(t)/\dT(t)$, where $\dA(t)$ and $\dT(t)$ are the axial and transversal diameters, respectively. The two silica beads have been reported in the figure to highlight the two areas where the forces are applied.
 Right panel: sketch of the deformation of the RBC in a linear shear flow with intensity $\dot{\gamma}$. \label{fig:rbc_stretch}}
\end{figure*}

%
\begin{figure}[t!]
\centering
\includegraphics[width=.9\linewidth]{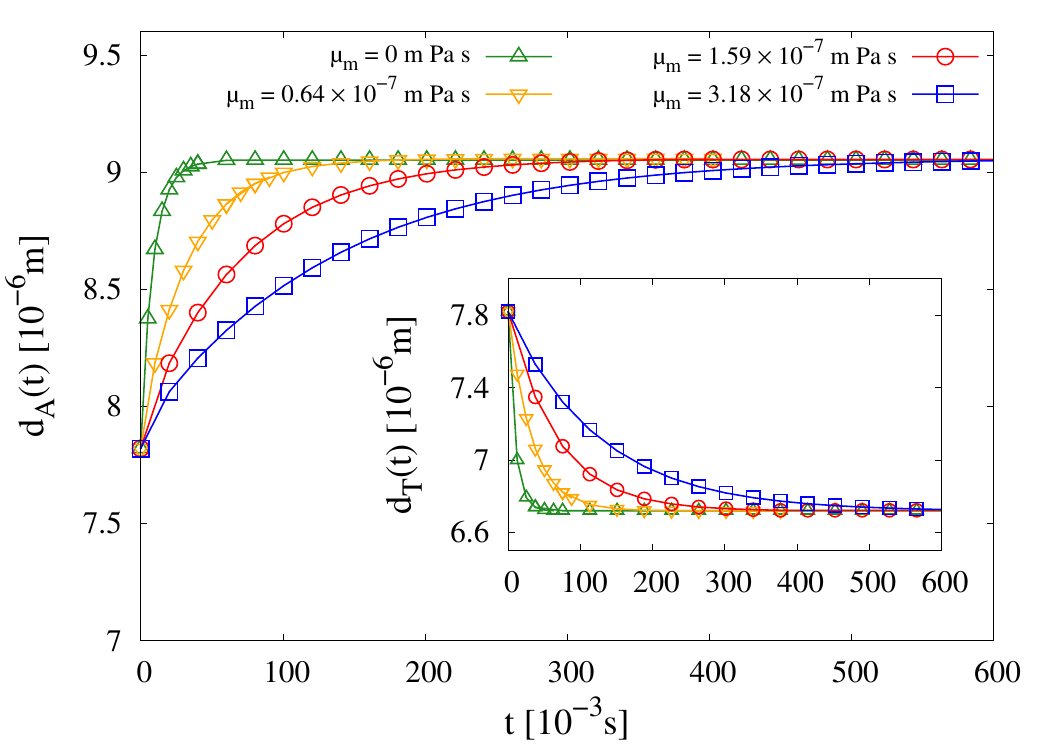}
\includegraphics[width=.9\linewidth]{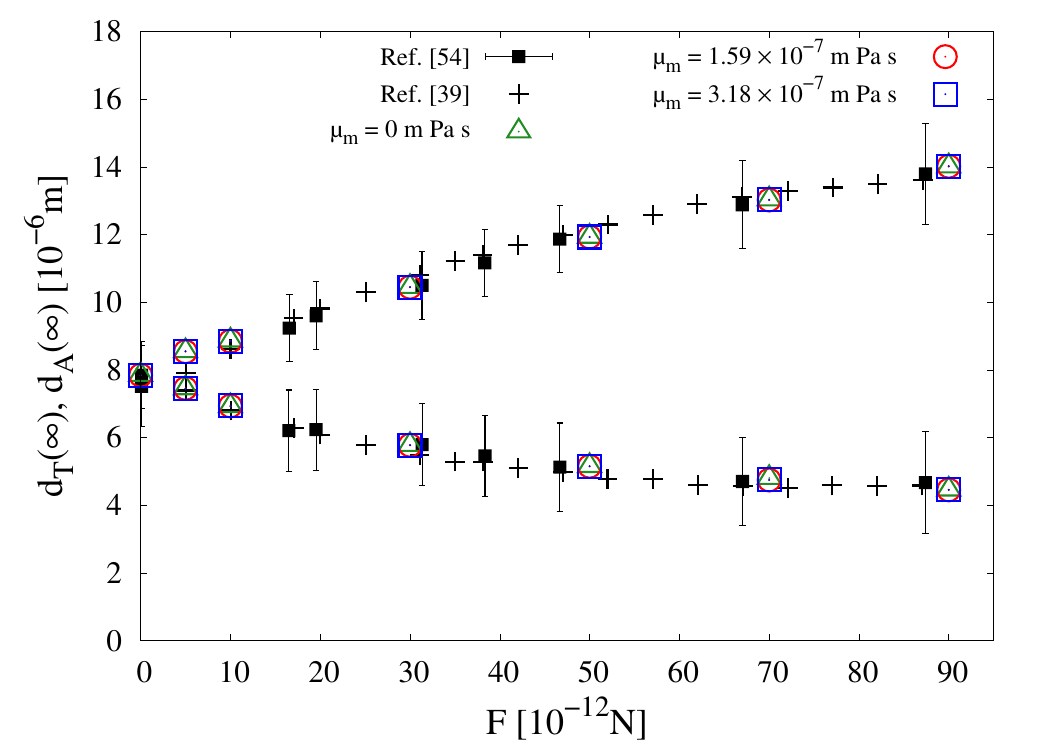}
\caption[figure]{Results for the stretching experiment of a single RBC with $F = 5\times 10^{-12}$~N (cfr. Fig.~\ref{fig:rbc_stretch}). Top panel: we report the axial ($\dA(t)$) and transversal ($\dT(t)$) diameters as a function of time $t$ for a given force $F=5\times 10^{-12}$ N and different values of the membrane viscosity $\mum$. Lines are guides for the eyes. Bottom panel: steady state values of the axial and transversal diameters as functions of the intensity of the applied force $F$ for different values of membrane viscosity $\mum$. Experimental data are also reported~\cite{art:suresh2005connections}, as well as numerical results~\cite{art:kruger14deformability}. A similar analysis for the RBC deformed in linear shear flow is presented in Fig.~\ref{fig:plot_shear}.
 \label{fig:plot_stretch}}
\end{figure}
%
\begin{figure}[th!]
\centering
\includegraphics[width=.9\linewidth]{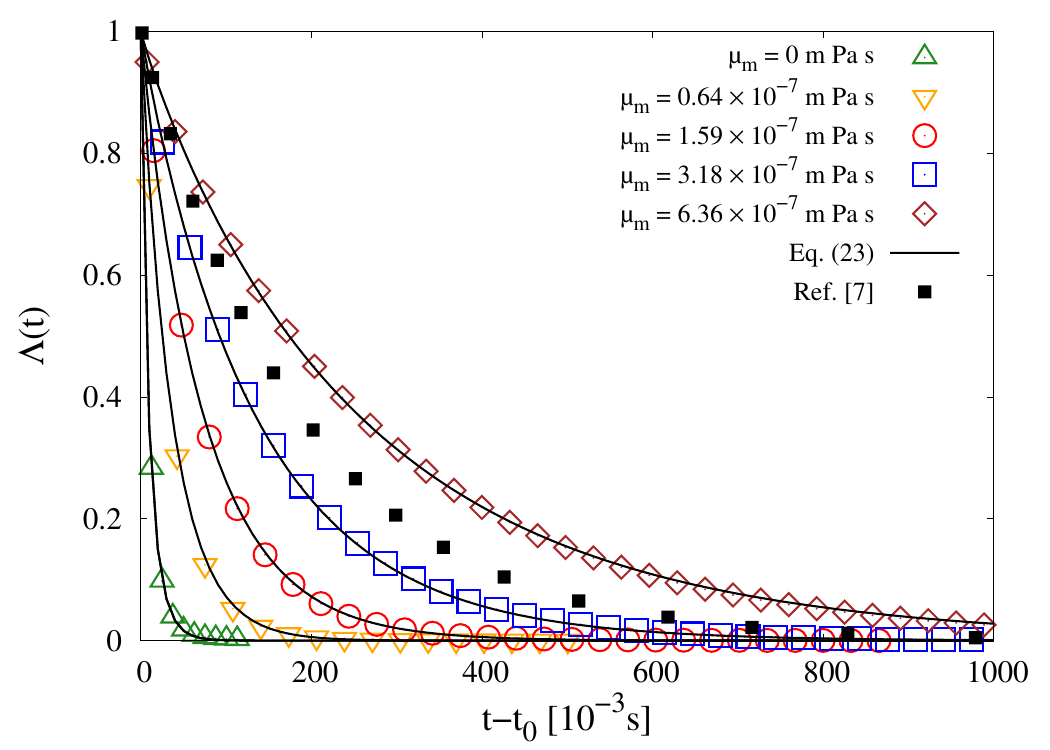}
\includegraphics[width=.9\linewidth]{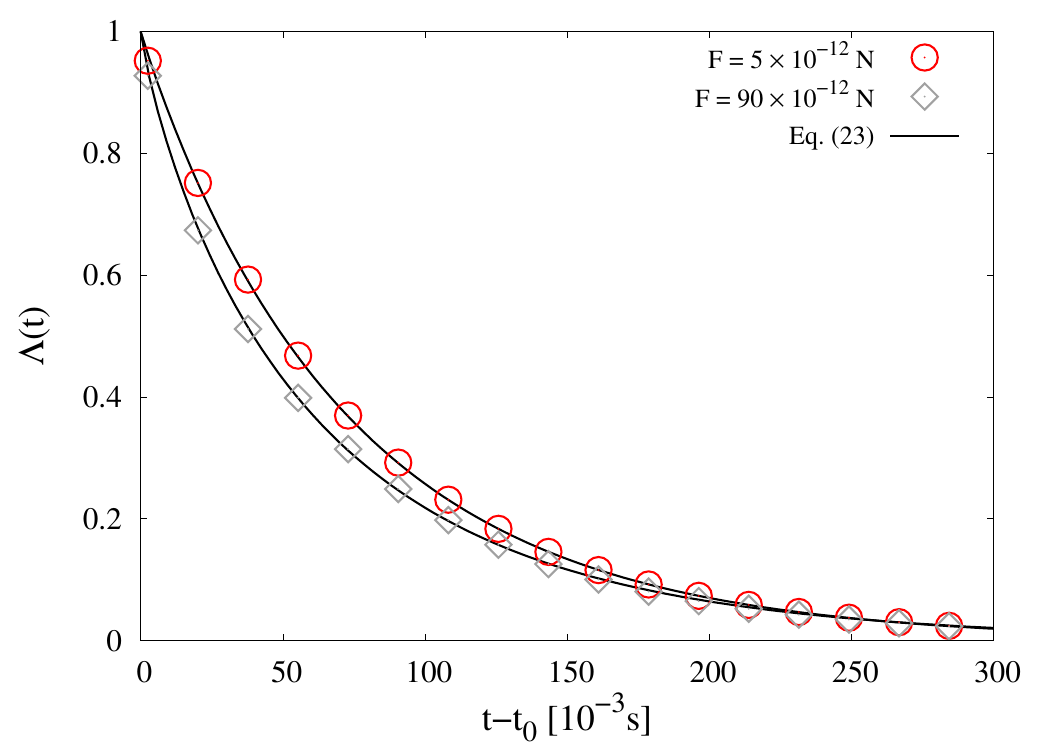}
\caption[figure]{RBC membrane relaxation after the sudden arrest of the mechanical load of intensity $F$. Top panel: we report the elongational index (cfr. Eq.~\eqref{eq:recovery}) as a function of time for a fixed $F=5\times 10^{-12}$~N and different values of membrane viscosity $\mum$. Experimental data are also reported~\cite{mills2004nonlinear}. Bottom panel: we report the elongational index as a function of time for a fixed value of membrane viscosity $\mum=1.59\times 10^{-7}$~m~Pa~s and different forces $F$. In both cases, the simulation data are fitted with a stretched exponential function (cfr. Eq.~\eqref{eq:tc_FIT}) from which we compute the relaxation time $t_c$. A similar analysis for the RBC deformed in linear shear flow is presented in Fig.~\ref{fig:shear_relax}. \label{fig:plot_relax}}
\end{figure}
\begin{figure}[th!]
\centering
\includegraphics[width=.9\linewidth]{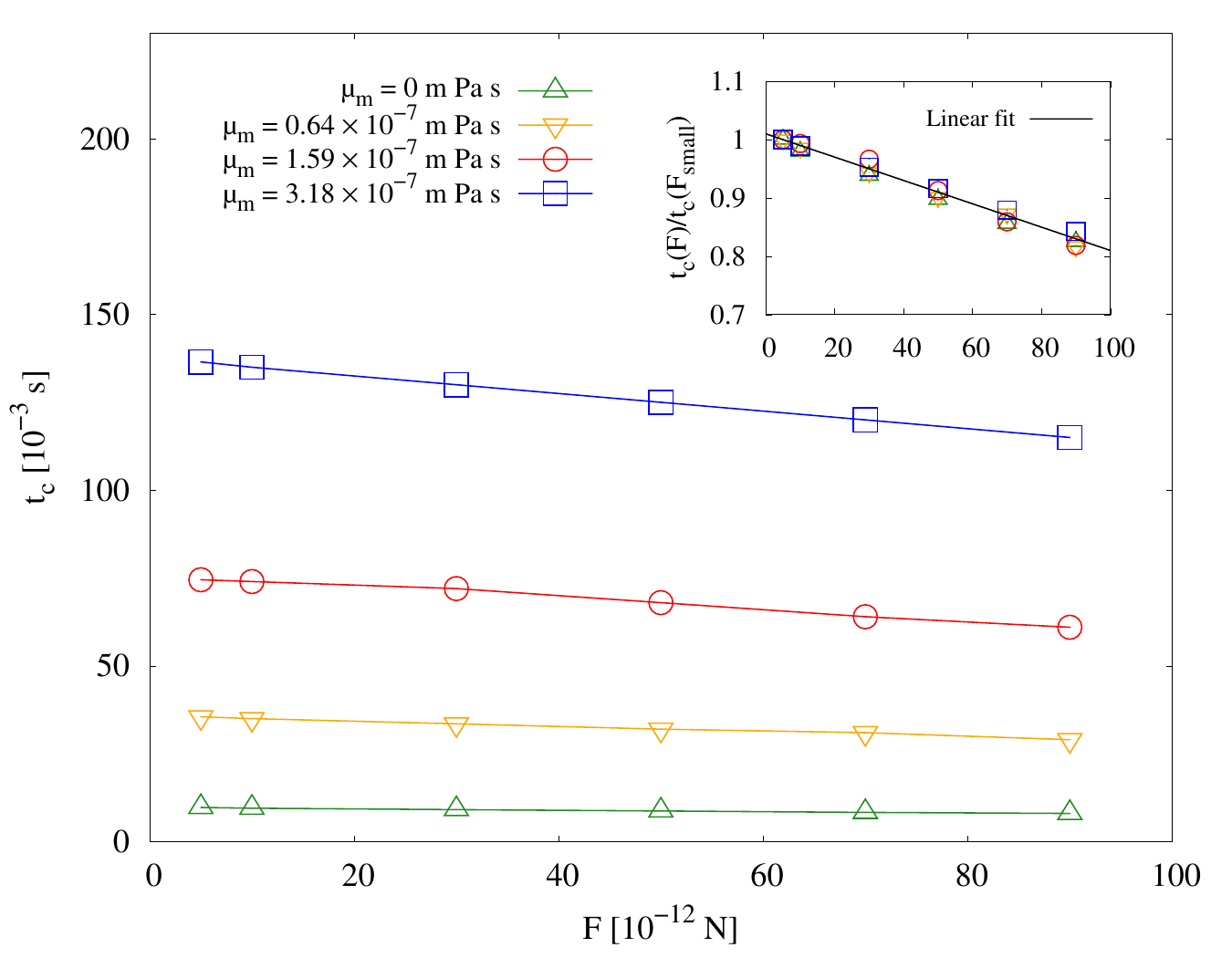}
\caption[figure]{RBC membrane relaxation after the sudden arrest of the mechanical load of intensity $F$. We report the relaxation time $t_c$ as a function of the forcing $F$, for different values of membrane viscosity $\mum$; lines are guides for the eyes. The inset shows normalised data with respect to the relaxation time at the smallest forcing analysed. Linear fit to the data is also provided (see text for details). \label{fig:plot_relax_force}}
\end{figure}

%
\section{Results on red blood cells (RBCs)}\label{sec:rbc_results}
RBC membrane comprises a cytoskeleton and a lipid bilayer, tethered together, forming a 2D thin shell~\cite{book:gommperschick,li2014erythrocyte}, characterised by a thickness of about three orders of magnitude smaller than the diameter. We recreate such a structure via a 3D triangular mesh with $\Nf\approx 4000$ faces, as depicted in Fig. \ref{fig:rbc}. To reconstruct it, we have employed the shape equation parametrised by~\cite{evans1972improved}:
\begin{equation}
z(x,y) = \pm \sqrt{1-\frac{x^2+y^2}{r^2}}\left(C_0+C_1\frac{x^2+y^2}{r^2}+C_2\left(\frac{x^2+y^2}{r^2}\right)^2\right)
\end{equation}
with $C_0 = 0.81\times 10^{-6} \mbox{ m}$, $C_1 = 7.83\times 10^{-6} \mbox{ m}$ and $C_2 = -4.39\times 10^{-6} \mbox{ m}$; $r=3.91\times 10^{-6} \mbox{ m}$ is the large radius (see Fig.~\ref{fig:rbc}). Regarding the elastic moduli (see Sec.~\ref{sec:membrane_model}), we use the bending modulus $\kB = 2\times 10^{-19}\mbox{ Nm}$~\cite{book:gommperschick}, the elastic shear modulus $\kS=5.3\times 10^{-6}\mbox{ Nm}^{-1}$~\cite{art:suresh2005connections} and the elastic dilatational modulus $\Kal=50 \kS$~\cite{thesis:kruger}. 
We chose the values of the elastic moduli in order to reproduce the dynamics of a young and healthy RBC: older erythrocytes are characterised by larger values of the shear modulus (about 20\%)~\cite{bronkhorst1995new}. Plasma viscosity is set to $\muout = 1.2\times 10^{-3}$~Pa~s, while the cytoplasm viscosity is $\muin = 6\times 10^{-3}$~Pa s~\cite{thesis:kruger}. All the aforementioned details are summarised in Fig.~\ref{fig:rbc}. A delicate issue pertains the (RBC) volume and (membrane) area conservation in the simulations. Following~\cite{thesis:kruger}, one may add to the model a surface energy with large surface modulus (imposing membrane area conservation) and a volume energy with a large volume modulus (imposing RBC volume conservation). In practice, we do not use such terms because we have verified that the incompressibility of the fluid and the local elastic energy are sufficient to retrieve excellent global area and volume conservation. Table~1 in the ESI section reports the maximum area and volume change experienced in each of the simulations with the RBCs that we will discuss below.\\
Considering the membrane viscosity, we put $\mus = \mud=\mum$~\cite{barthes1985}, with the membrane viscosity varying in the range $\mum\in[0, 3.18\times 10^{-7}]$~m~Pa~s (note that $\mum$ is the viscosity of the 2D membrane and therefore it is measured in [m Pa s]). The value of the artificial spring in the SLS model is  $k' = \frac{100}{3}\kS$ (see Sec.~\ref{sec:membrane_model}).
We wish to stress that, differently from Section~\ref{sec:benchmarks}, we directly use $\mum$ and $\dot{\gamma}$ instead of $\Bq$ and $\Ca$ to quantify the effects of membrane viscosity and the intensity of the shear flow. These can be translated in $\Bq$ and $\Ca$ once a characteristic length scale is chosen. For example, if we choose the main radius $r=3.91\times 10^{-6}$ m as characteristic length scale to compute both $\Bq$ (cfr. Eq.~\eqref{eq:bq_capsule}) and $\Ca$ (cfr. Eq.~\eqref{eq:ca_capsule}), we obtain  $\Bq = \frac{\mum}{\muout r} \in [0,50]$ and $\Ca = \dot{\gamma}\frac{\muout r}{\kS} \in[0.01,0.10]$, respectively.\\
To study the transient dynamics of the RBC, we characterised the relaxation time $t_c$ following the sudden arrest of an external mechanical load~\cite{fedosov2010multiscale,evans1976membrane,chien1978theoretical,braunmuller2012hydrodynamic,tran1984determination,riquelme2000determination}. Going back to the general motivations that we have exposed in Section~\ref{sec:intro}, we stress once more that the characterisation of such a relaxation time provides important insights on the choice of the parameters in the so-called ``tensor based'' hemolysis models for VADs design and improvement~\cite{art:arora04,art:arora2006hemolysis,art:pauli2013transient,art:gesenhues2016strain}. 
Starting from the steady, deformed erythrocyte configurations, we have characterised the shape recovery by suddenly switching off the external mechanical load and by monitoring the extension ratio $\lambda(t) = \dA(t)/\dT(t)$, where $\dA(t)$ and $\dT(t)$ are the axial and transverse diameters, respectively (see Fig.~\ref{fig:rbc_stretch}). We then constructed the time-dependent elongational index~\cite{hochmuth1979red,fedosov2010multiscale}
\begin{equation}\label{eq:recovery}
\Lambda(t)=\frac{[\lambda(t)-\lambdafin][\lambdain+\lambdafin]} {[\lambda(t)+\lambdafin][\lambdain-\lambdafin]}
\end{equation}
where $\lambdain$ and $\lambdafin$ correspond to the values of the extension ratio at the beginning and the end of the relaxation process, respectively. The elongational index is constructed in such a way that it has an initial value of 1 and tends to 0 when the original shape is recovered. Following Fedosov~\cite{fedosov2010multiscale} we obtain the relaxation time $t_c$ via a fit through the stretched exponential
\begin{equation}\label{eq:tc_FIT}
\Lambda(t)=\exp\left\{-\left(\frac{t}{t_c}\right)^\delta\right\}.
\end{equation}
Some considerations on the use of the stretched exponential function instead of a single exponential function will be brought forward when analysing data.
In the following, we show the numerical results on the deformation and the shape recovery of a single RBC deformed by an external mechanical load. In Sec.~\ref{sec:stretching}, we numerically study the RBC deformation caused by two forces applied at the RBC ends, a set-up that is meant to mimic the deformation of RBC experimentally observed in optical tweezers~\cite{art:suresh2005connections}. In Sec.~\ref{sec:shear} we study the RBC deformation and shape recovery under the imposition of a given shear rate. In Sec.~\ref{sec:comparison} we will compare the results from both protocols. The membrane viscosity $\mum$ and the load strength will be kept as variable input parameters.

\subsection{Stretching}\label{sec:stretching}
We simulate the stretching of a single erythrocyte induced by two constant forces of equal intensity $F=|\vec{F}|$ and opposite direction, applied at the ends of the RBC (see Fig.~\ref{fig:rbc_stretch}, left panel). Simulations are performed in a 3D box $\Lx \times \Ly \times \Lz = (28 \times 12 \times 12)\times 10^{-6}~\mbox{ m}$, with full periodic boundary conditions along all directions; we study the cell deformation by varying both the membrane viscosity $\mum$ and forcing intensity $F$. The set-up that we use is meant to mimic the stretching experiments conducted on a single RBC in an optical tweezer, wherein two silica beads (whose diameter is $d_s=4.12\times 10^{-6}$ m) are attached to the two ends of the erythrocyte~\cite{art:suresh2005connections}. In the simulations, we consider two areas at the end of the membrane which are compatible with the dimensions of the silica beads used in the experiments~\cite{art:suresh2005connections}, and we apply a force $\tilde{\vec{F}}=\vec{F}/n_s$ at each of the $n_s$ nodes that lie in this area. Notice that previous numerical studies with the IB-LBM by Kr\"{u}ger {\it et al.}~\cite{art:kruger14deformability} already investigated the deformation of a single RBC under the set-up that we use. Here we want to further extend such a study by highlighting the role of membrane viscosity. In the top panel of Fig.~\ref{fig:plot_stretch}, we show the transient dynamics of both $\dA(t)$ and $\dT(t)$ in the presence of a force intensity $F=5\times 10^{-12}$ N, for different values of $\mum$. We observe that the membrane viscosity plays a paramount role in the approach towards the steady state (in fact, a larger $\mum$ corresponds to larger time needed to reach the steady state), while it does not affect the value of such a steady state deformation (bottom panel). This is quite apparent, in view of the fact that there is no stationary flow for large times, hence dissipation on the membrane does not take place. Results on the relaxation time are reported in Fig.~\ref{fig:plot_relax}. In the top panel, we report the time evolution of the elongational index $\Lambda(t)$ (cfr. Eq.~\eqref{eq:recovery}) after the arrest of an applied forcing $F=5\times 10^{-12}$ N, for different values of $\mum$. We chose $F=5\times 10^{-12}$ N (that is the smallest force that we simulated) because we expect that forcing effects are negligible for such a low value of $F$, hence the analysis highlights the intrinsic properties of the viscoelastic membrane. But again, the increase of $\mum$ results in slower recovery dynamics. We have reported experimental data from Mills~\textit{et al.}~\cite{mills2004nonlinear}. Only for this case, we performed a simulation with a value of membrane viscosity $\mum = 6.36\times 10^{-7}$~m~Pa~s, just to show that the simulated range of membrane viscosities is a realistic one. We further studied the dependency of the relaxation time on the applied forcing strength. Results are reported in the bottom panel of Fig.~\ref{fig:plot_relax}. Again, we plot the evolution of the elongational index $\Lambda(t)$, but now we fix the membrane viscosity $\mum=1.59\times 10^{-7}$ m Pa s and we vary the force strength $F$ by an order of magnitude. Larger forcing (i.e. more deformed shape) results in faster recovery dynamics. In both cases, we observe that Eq.~\eqref{eq:tc_FIT} provides an excellent fit to the data. Thus, the relaxation time $t_c$ can be accurately retrieved: the values of $t_c$ as a function of $F$ for different values of membrane viscosity $\mum$ are reported in Fig.~\ref{fig:plot_relax_force}, highlighting that the relaxation time changes at least of a factor 15-20 \% when the forcing increases. More quantitatively, for a given value of $\mum$, we have analysed data by rescaling the relaxation time with the value corresponding to the smallest forcing available, i.e. $\fsmall =5\times 10^{-12}\mbox{N}$: results are displayed in the inset of Fig.~\ref{fig:plot_relax_force}. Remarkably, data for different $\mum$ collapse together on a master curve, that we linearly fit with $t_c(F)/t_c(\fsmall) = a_0 - a_1 \, F$, with $a_0=1.01$ and $a_1=2\times 10^{9}$ N$^{-1}$.  This collapse shows that the functional dependency on the forcing can be factorised with respect to the dependency on the membrane viscosity. 
Before closing this section, some remarks on the use of the stretched exponential function Eq.~\eqref{eq:tc_FIT} are needed. From one side, as already pointed out in~\cite{fedosov2010multiscale}, the use of $\delta \neq 1$ is instrumental to improve the quality of the fit of the relaxation dynamics of $\Lambda(t)$. All the values of $\delta$ have been summarised in Fig.~5 of the ESI section: $\delta$ is close to 1, and predominantly located in the range 0.9-1.0. A value of $\delta \neq 1$ could be indicative of the presence of multiple time scales in the relaxation process. To delve deeper into this issue, in the ESI section, we provide log-lin plots of all the data analysed in Fig.~\ref{fig:plot_relax}. We find that slower time scales clearly appear below (or even well below) $\Lambda(t) \sim 0.01$, while the stretched exponential fit works very well in the range $0.01 \le \Lambda(t) \le 1$. The crossover time scale as well as the details of the late dynamics are found to depend both on the membrane viscosity as well as the load strengths. In particular, the larger is the membrane viscosity and the later the slower dynamics takes place; the larger is the forcing and the slower is the late dynamics. We hasten to remark, however, that the dynamics at such very small $\Lambda(t)$ could be well affected by numerical discretisation artefacts, hence a dedicated study by its own would be necessary to make more quantitative claims.

%
\subsection{Linear shear flow}\label{sec:shear}
%
\begin{figure}[t!]
\centering
\includegraphics[width=.9\linewidth]{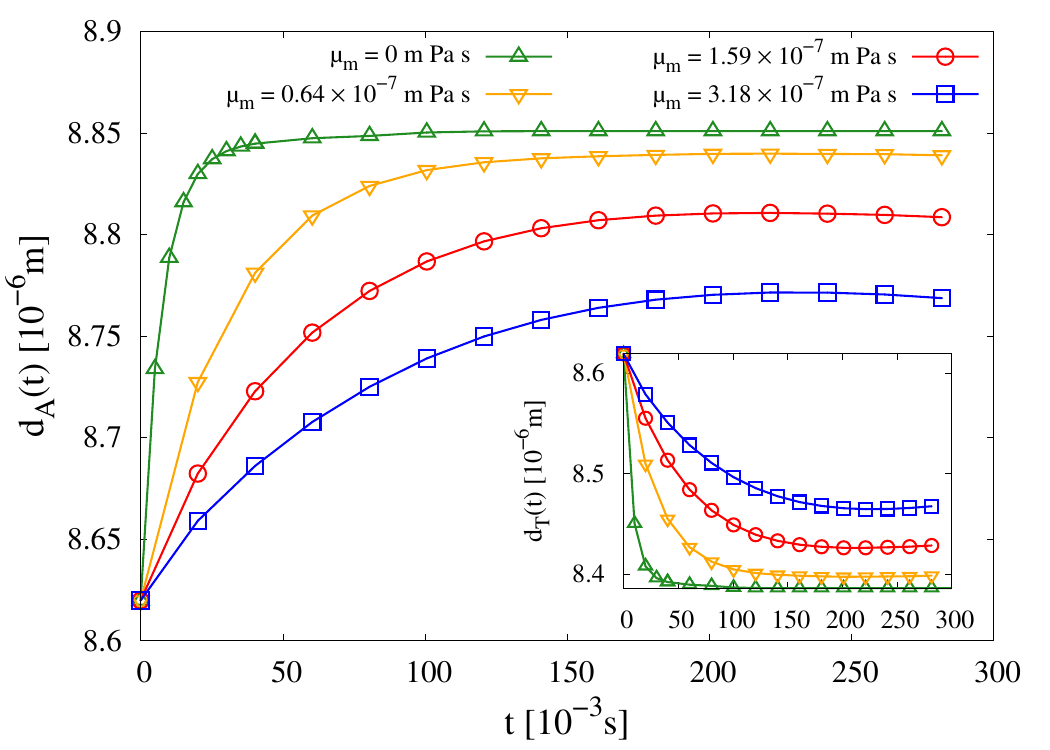}
\includegraphics[width=.9\linewidth]{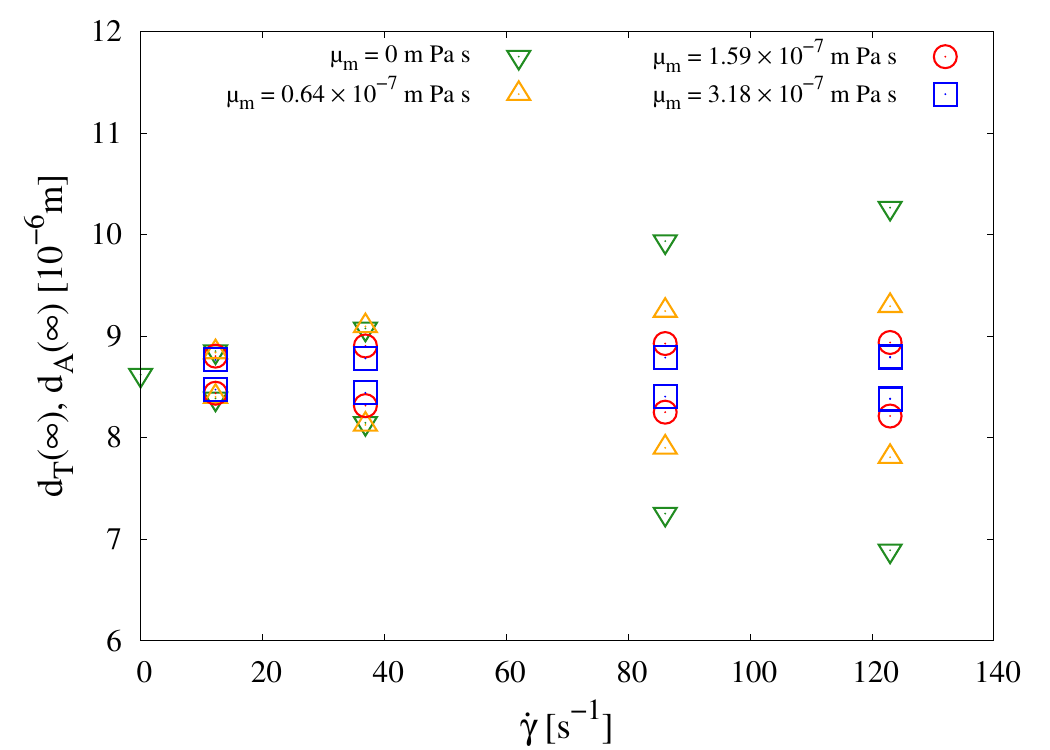}
\caption[figure]{Results for the deformation of a single RBC in linear shear flow with intensity $\dot{\gamma}$ (cfr. Fig.~\ref{fig:rbc_stretch}). Top panel: we report the axial ($\dA(t)$) and transversal ($\dT(t)$) diameters as a function of time $t$ for a given shear rate $\dot{\gamma}=1.23 s^{-1}$ and different values of the membrane viscosity $\mum$. Lines are guides for the eyes. Bottom panel: steady state values of the axial and transversal diameters as functions of the intensity of the applied shear rate $\dot{\gamma}$ for different values of $\mum$. \label{fig:plot_shear}}
\end{figure}
\begin{figure}[t!]
\centering
\includegraphics[width=.9\linewidth]{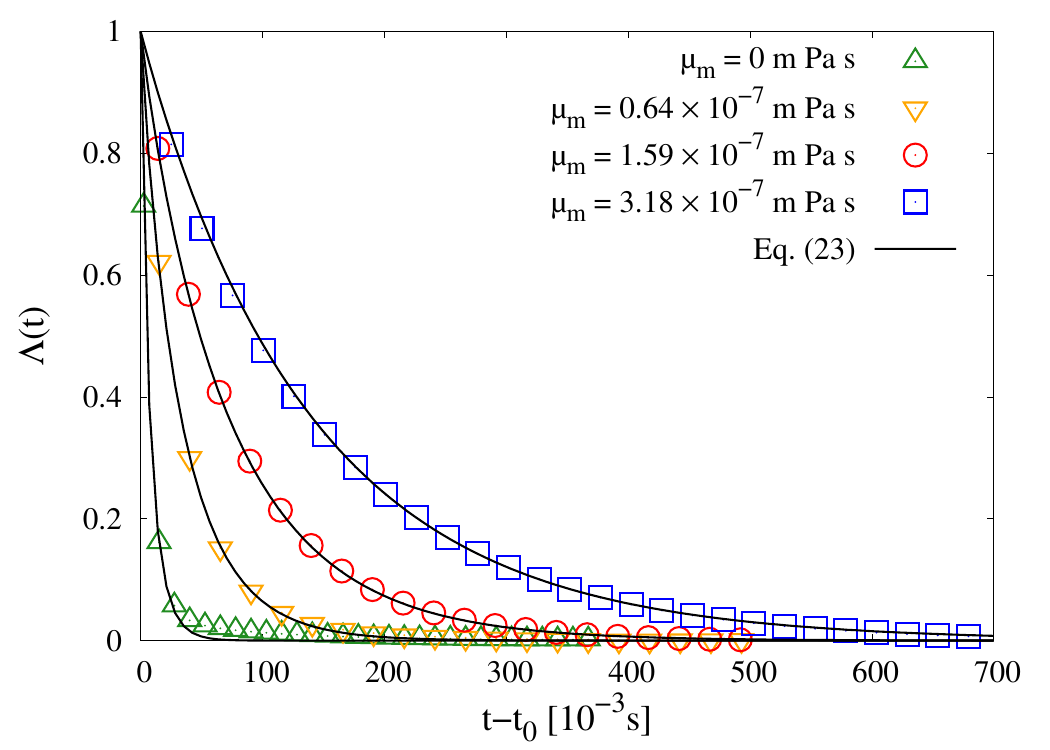}
\includegraphics[width=.9\linewidth]{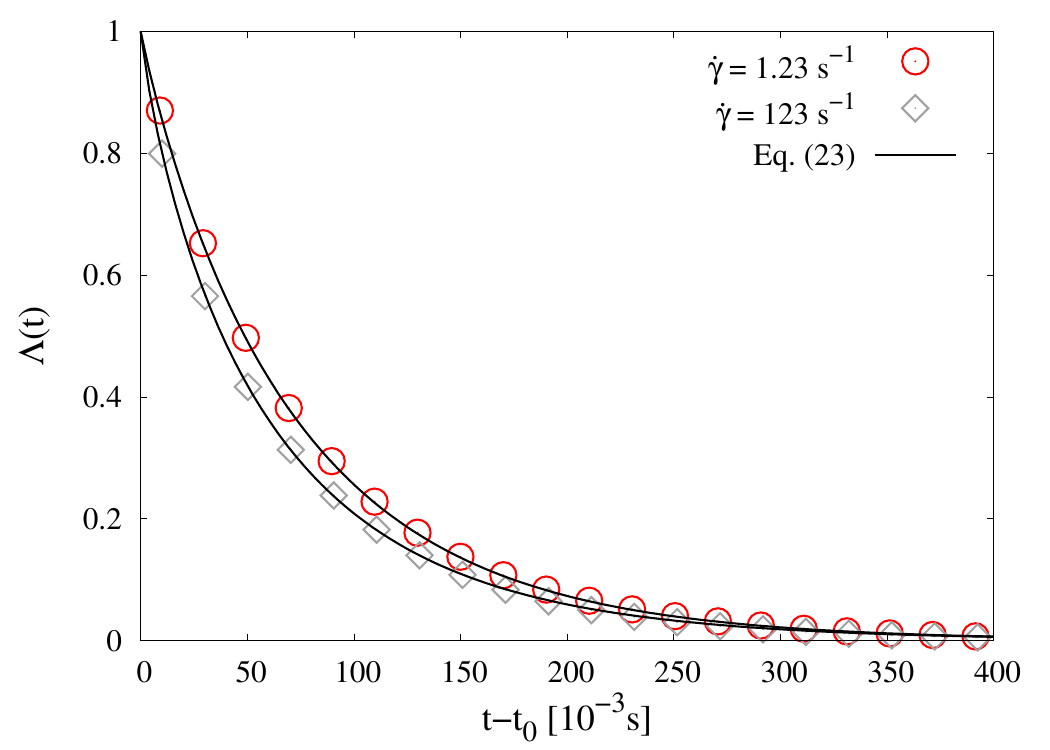}
\caption[figure]{RBC membrane relaxation after the sudden arrest of the shear flow of intensity $\dot{\gamma}$. Top panel: we report the elongational index (cfr. Eq.~\eqref{eq:recovery}) as a function of time for a fixed  $\dot{\gamma}=1.23\ s^{-1}$ and different values of membrane viscosity $\mum$. Bottom panel: we report the elongational index as a function of time for a fixed value of membrane viscosity $\mum=1.59\times 10^{-7}$~m~Pa~s and different shear rates $\dot{\gamma}$. In both cases, the simulation data are fitted with a stretched exponential function (cfr. Eq.~\eqref{eq:tc_FIT}) from which we compute the relaxation time $t_c$. \label{fig:shear_relax}} 
\end{figure}
\begin{figure}[t!]
\includegraphics[width=.9\linewidth]{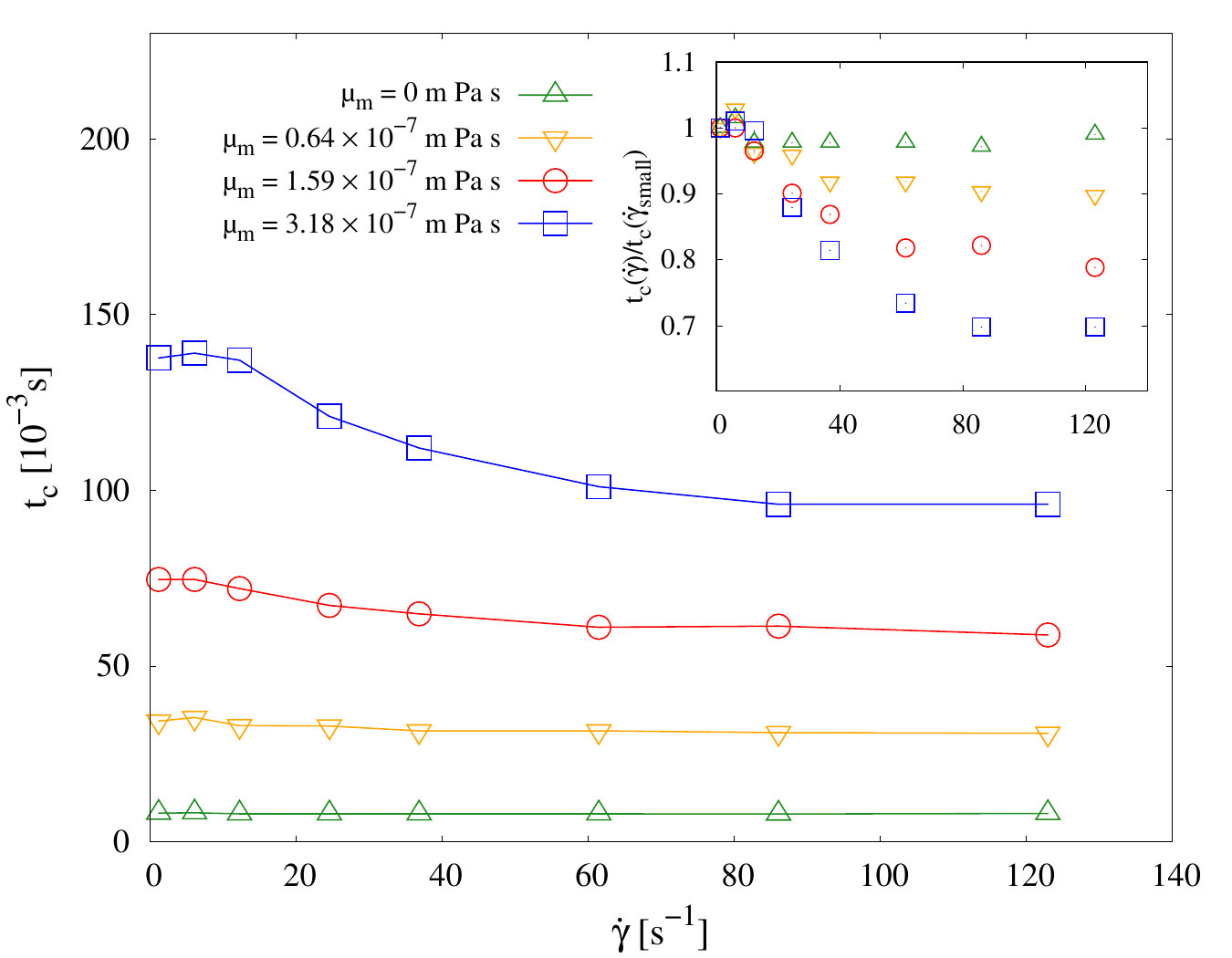}
\caption[figure]{RBC membrane relaxation after the sudden arrest of the shear flow of intensity $\dot{\gamma}$. We report the relaxation time $t_c$ as a function of the shear rate $\dot{\gamma}$, for different values of membrane viscosity $\mum$; lines are guides for the eyes. 
\label{fig:shear_time}}
\end{figure}
%

\begin{figure*}[t!]
\centering
\includegraphics[width=.47\linewidth]{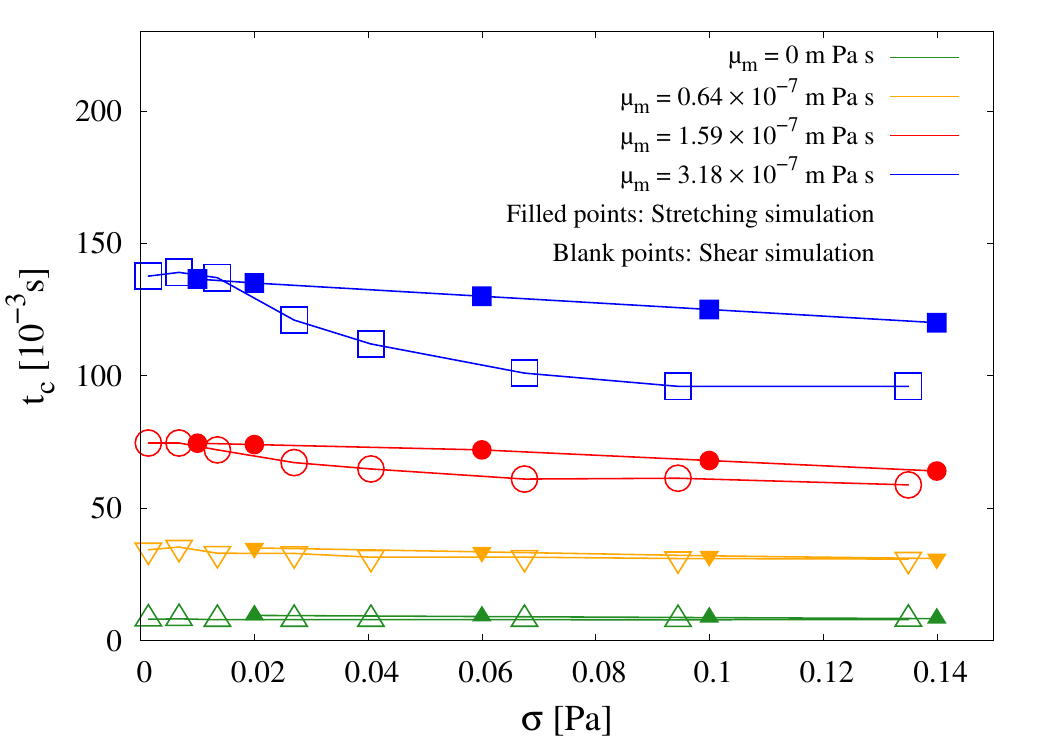}
\includegraphics[width=.47\linewidth]{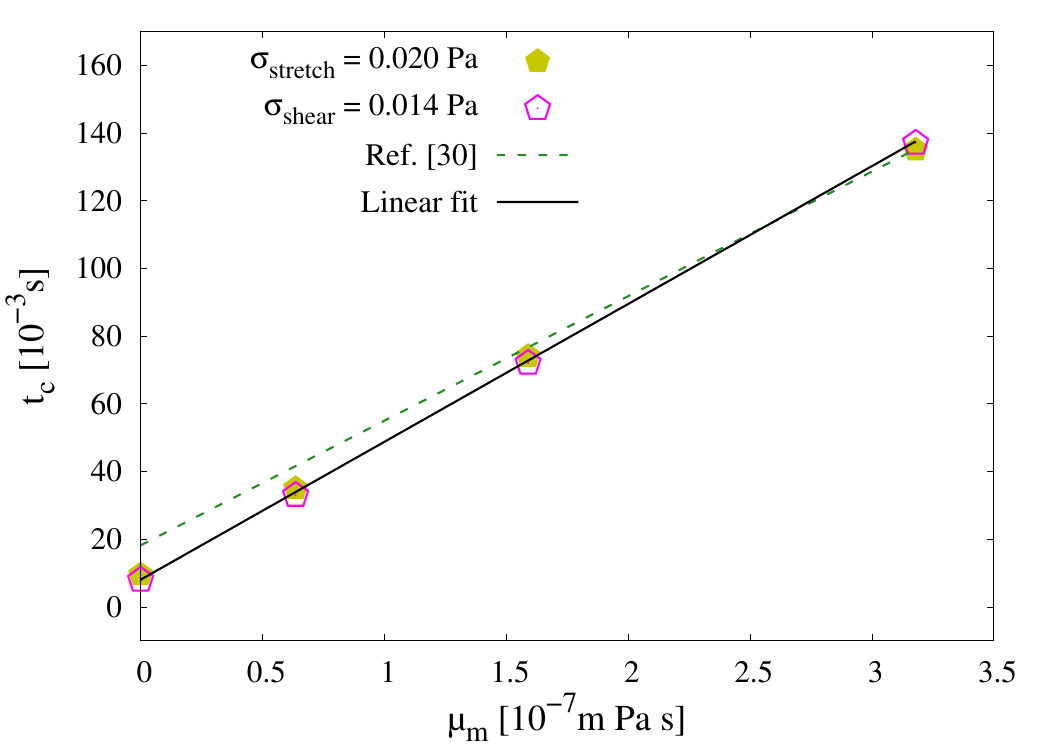}
\caption[figure]{Comparison between stretching (filled points) and shear (blank points) experiments. Left panel: we report the relaxation time $t_c$ as a function of the mechanical load strength $\ml$ for different values of membrane viscosity. Details on the computation of the load strength are given in the text. Right panel: we show the relaxation time as a function of the membrane viscosity for small representative load strength simulated in both the stretching and shear experiment. Results are compared with the phenomenological estimate in~\cite{art:prado15}. Linear fit to the data is also provided. \label{fig:tc_mech_load}}
\end{figure*}

We simulate the deformation of a single RBC induced by an imposed shear with intensity $\dot{\gamma}$ (that is the same setup used in benchmark section, see Sec.~\ref{sec:benchmarks}). This load mechanism differs from the stretching force considered in the previous section, hence we can investigate the dependency of the relaxation time on the typology of the mechanical load. Simulations are performed in a 3D box $\Lx \times \Ly \times \Lz = (20 \times 20 \times 32)\times 10^{-6}~\mbox{ m}$ with periodic boundary conditions set along $x$ and $y$ directions. As for the droplet and spherical capsule case (cfr. Sec.~\ref{sec:benchmarks}), we set the wall velocity $\Uw=(\pm \dot{\gamma}\Lz/2,0,0)$ on the two plane walls at $z=\pm \Lz/2$ ($\Uw>0$ for $z>0$), while periodic boundary conditions are set along $x$ and $y$ directions. The orientation of the RBC with respect to the shear flow is shown in Fig.~\ref{fig:rbc_stretch}. 
Since we are interested in studying an alternative mechanical load to be compared with the stretching experiment in optical tweezers, we decided to place the major axial diameters of the RBC in the shear plane. Indeed, for the range of $\dot{\gamma}$ that we simulate, rolling dynamics is expected~\cite{mauer2018flow}; hence the chosen orientation is somehow more appropriate to avoid complications coming from the emergence of time-dependent motion~\cite{minetti2019dynamics,cordasco2017shape}. The 3D mesh used to represent the RBC is the same used in the stretching experiment. Now, we vary both the membrane viscosity $\mum$ and the intensity of shear rate $\dot{\gamma}$. Similarly to what we have done for the stretching experiment (see Fig.~\ref{fig:plot_stretch}), we first look at the dynamics of the deformation of the RBC under a linear shear flow and at the steady state shape. In the top panel of Fig.~\ref{fig:plot_shear}, we plot the transient dynamics of main diameters $\dA(t)$ and $\dT(t)$. Similarly to the stretching experiment, we found a slower dynamics when the membrane viscosity $\mum$ increases; on the other hand, unlike the stretching experiment, the steady deformation is now affected by the membrane viscosity because the steady states have a non zero velocity acting on the surface of the RBC. The results on the steady state deformation are reported in the bottom panel of Fig.~\ref{fig:plot_shear}. Increasing the membrane viscosity results in a smaller deformation, which is in line with what we found in Section~\ref{sec:bench_membrane_viscosity} (see Fig.~\ref{fig:bench_droplet}). Moreover, for the largest values of 
$\dot{\gamma}$, we noticed that the dynamics of the elongational index shows oscillations around a mean value when approaching the steady state (data not shown), similarly to what happens for a viscous capsule (see Fig.~\ref{fig:li_fig7}). For a fixed shear rate $\dot{\gamma}$ and membrane viscosity $\mum$, these oscillations go to zero in time. Again, to further investigate the transient dynamics of the RBC after the arrest of the shear flow, we look at the time-dependent elongational index $\Lambda(t)$ (Eq.~\eqref{eq:recovery}) and we use Eq.~\eqref{eq:tc_FIT} to fit our data. In the top panel of Fig.~\ref{fig:shear_relax}, we report the time evolution of $\Lambda(t)$ with constant shear rate $\dot{\gamma} = 1.23\mbox{ s}^{-1}$ for different values of membrane viscosity $\mum$. As already observed in the stretching experiment (see Fig.~\ref{fig:plot_relax}), increasing the membrane viscosity $\mum$ results in a slower recovery dynamics. In the bottom panel of Fig.~\ref{fig:shear_relax}, we study the effect of the shear rate $\dot{\gamma}$ for a fixed value of membrane viscosity $\mum = 1.59\times 10^{-7}$~m~Pa~s. Again, a higher strength of the mechanical load (i.e. higher shear rate $\dot{\gamma}$) corresponds to a lower relaxation time. Finally, all the values of $t_c$ that we extracted based on the Eq.~\eqref{eq:tc_FIT} are reported in Fig.~\ref{fig:shear_time} for different values of $\dot{\gamma}$ and $\mum$. 
In contrast to the stretching experiment, we notice that for fixed membrane viscosity the relation between $t_c$ and $\dot{\gamma}$ does not show a net linear behaviour; rather, for the largest $\mum$ we observe the emergence of some non-linear behaviour which is steeper at smaller $\dot{\gamma}$. Some important additional information is also conveyed by the inset in Fig.~\ref{fig:shear_time}, where we report data normalised with respect to the relaxation time at the smallest considered $\dot{\gamma}$, i.e. $\dotgammasmall = 1.23\mbox{ s}^{-1}$. We see that the dependence of the relaxation time on the shear rate is triggered by a non-zero membrane viscosity: while data at $\mum = 0$ do not show an appreciable shear rate-dependency, when $\mum > 0$ such a dependency emerges and it gets more apparent at increasing~$\mum$. Regarding the values of $\delta$ for the shear case, these fall in the same ballpark of those observed for the stretching experiment (see Fig.~5 in the ESI section). Log-lin plots of data in Fig.~\ref{fig:shear_relax} (see Fig.~4 in the ESI section) actually show that the late-time dynamics for these experiments does not necessarily show slower dynamics, as for the case of the stretching experiment. Again, we remark that effects taking place at very small values of $\Lambda(t)$ need dedicated higher resolution investigations to be more quantitatively assessed.
\begin{table*}[ht!]
\centering
 \begin{tabular}{llll}  
 \multicolumn{4}{c}{} \\
\hline
 \qquad \qquad Author & \qquad $t_c$ [s] & $\mum$ [$10^{-7}$ m Pa s] &  \qquad Technique \\
Evans \& Hochmuth 1976~\cite{evans1976membrane} & 0.300 & $\sim 1$ & Micropipette aspiration\\
Chien {\it et al.} 1978~\cite{chien1978theoretical}  & $0.146 \pm 0.055$ & 0.6 - 4.0 & Micropipette aspiration \\
Hochmuth {\it et al.} 1979~\cite{art:hochmuth79}& 0.100 - 0.130 & 6 - 8 & Micropipette aspiration \\
Tran-Son-Tay {\it et al.} 1984~\cite{tran1984determination}  & - & $ 0.53 - 0.96$ & Tank-treading \\
Baskurt \& Meiselman 1996~\cite{art:baskurt96}  & $0.119 \pm 0.017$ & - & Shear (light reflection) \\
Baskurt \& Meiselman 1996~\cite{art:baskurt96}  & $0.097 \pm 0.015$ & - & Shear (ektacytometry) \\
Riquelme {\it et al.} 2000~\cite{riquelme2000determination}  & - & $  2.7 - 4.1 $ & Sinusoidal shear stress\\
Tomaiuolo \& Guido 2011~\cite{art:tomaiuolo11}  & 0.100 & 4.7 - 10.0 & Microchannel Deformation \\  
Braunm\"{u}ller {\it et al.} 2012~\cite{braunmuller2012hydrodynamic}  & 0.100 - 0.130 & $\sim$10 &  Micropipette aspiration \\
Prado {\it et al.} 2015~\cite{art:prado15}  & $0.08\pm 0.01$ & $  0.6-0.9 $ & Numerical and experimental\\
Fedosov 2010\cite{fedosov2010multiscale} & 0.100 - 0.130 & $\sim 1^{\dagger}$ & Numerical simulation \\
Present work & 0.100 - 0.140 & $\sim 3.18$ & Numerical simulation \\
\hline
\end{tabular}
\caption{Experimental and numerical relaxation times.\\
$^{\dagger}$ Fedosov reports the 3D membrane viscosity $\mum^{3D} = 0.022$~Pa~s. We multiplied this value by the radius of the RBC to get $\mum$.}
\label{tab:relaxation_times}
\end{table*}

\subsection{Comparison between load mechanisms}\label{sec:comparison}
Given the analysis for the relaxation times that we have performed in Secs.~\ref{sec:stretching}-\ref{sec:shear}, the natural next step is the quantitative comparison between the two load mechanisms. In order to make the comparison fair, we need to appropriately compare the strengths of the two load mechanisms. To do so, we decided to look at the total stress $\sigmat$ of the system in the initial configuration (prior to the relaxation process). The total stress can be split in two contributions, one coming from the particle ($\sigmap$) and one from the fluid ($\sigmaf$)~\cite{thesis:kruger,gross2014rheology}. In the stretching experiment there is no net flow, hence the particle stress contributes to the total stress; moreover, when we stretch the RBC with two opposite forces along the $x$ axis, the main contribution to $\sigmap$ comes from $\sigmap_{\mbox{\tiny xx}}$ (the other diagonal components are smaller but not zero, $\sigmap_{\mbox{\tiny yy}}\approx\sigmap_{\mbox{\tiny zz}}<\sigmap_{\mbox{\tiny xx}}$). In the shear experiment, instead, the major contribution to the total stress comes from the fluid part and the stress tensor is dominated by the off-diagonal xz-component, while diagonal components are non zero but smaller in magnitude. Given these facts, we decided to study the relaxation times as a function of the value of the dominant total stress contribution (hereafter load strength), i.e. $\sigmat_{\mbox{\tiny xx}}$ for the stretching experiment and $\sigmat_{\mbox{\tiny xz}}$ for the shear experiment. Results are reported in Fig.~\ref{fig:tc_mech_load}. In the Left panel, we report the results of Fig.~\ref{fig:plot_relax_force} (filled points) and Fig.~\ref{fig:shear_time} (blank points) as a function of the corresponding load strength. Results do not overlap exactly, and this is not surprising, since the two load mechanisms are actually different in nature: while in the stretching experiment, the erythrocyte starts the relaxation process with the fluid at rest, in the shear experiment the RBC is immersed in a fluid flow with a given velocity profile. This reflects on the different stress contributions that we measure in the two cases: while in the stretching experiment the main stress component is related to the diagonal (xx) component, in the shear experiment the main contributions are off-diagonal (i.e. xz). Thus, it is not surprising that the results of the two simulations do not show a precise correspondence. Nevertheless, one has to notice that, at fixed membrane viscosity, the two sets of data are somehow comparable in the order of magnitude. In other words, the value of the membrane viscosity sets the order of magnitude of the relaxation time, although the detailed functional dependency is affected by the chosen load mechanism. Large deformations imply smaller $t_c$: in this regime there may be non-linear forcing effects as well as multiple time scales to come into play. Notice, indeed, that for the largest membrane viscosity used, the $\delta$ parameter in Eq.~\eqref{eq:tc_FIT} shows a decreasing trend with the applied forcing (see Fig.~5 in the ESI section).\\
Independently of the load mechanism, however, one expects that when the load strength goes to zero the relaxation time should not depend on the chosen load mechanism: in such a limit, in fact, we should recover the membrane ``intrinsic'' relaxation time. This is indeed apparent from our data, as they show a neat asymptotic behaviour in the limit of vanishing load strength. 
Such a trend is better highlighted in the right panel of Fig.~\ref{fig:tc_mech_load}, in which we report the relaxation time $t_c$ as a function of the membrane viscosity $\mum$, both for the stretching and shear experiments, for the smallest values of simulated mechanical loads. This intrinsic relaxation time is in the order of few ms when $\mum$ goes to zero, which well agrees with the expectations in absence of membrane viscosity~\cite{evans19891}; then, as $\mum$ increases, the  intrinsic relaxation time linearly becomes of the order of hundreds of ms, which, again, is in good agreement with the values that can be retrieved from the experiments~\cite{art:hochmuth79,art:baskurt96,art:tomaiuolo11,chien1978theoretical,braunmuller2012hydrodynamic,art:prado15} (see Tab.~\ref{tab:relaxation_times}). The linear fit $t_c=a_0+a_1 \mum$, with $a_0=10.08\times 10^{-3}$ s and 
$a_1=0.395\times 10^{4}$ $\mbox{m }\mbox{N}^{-1}$, 
well fits the data. We also plot the linear prediction by Prado {\it et al.}~\cite{art:prado15} which differs by roughly 10\% by our best fit.

\section{Conclusions}\label{sec:conclusions}
Computational Fluid Dynamics (CFD) is widely used to achieve optimal design of biomedical equipments, such as Ventricular Assist Devices (VADs)~\cite{art:behbahani09,murakami1979nonpulsatile,nonaka2001development,art:anderson00,art:miyazoe98,art:yu00,art:qian00,art:allaire99,art:burgreen01}. Blood circulation in these mechanical pumps is simulated via a multi-scale approach involving transient dynamics of red blood cells (RBCs) at mesoscales~\cite{art:arora04}. The quantitative information on such transient dynamics is typically retrieved from experiments, which obviously do not adapt to the variety of conditions which may be encountered in the pump. In this paper, we have taken an alternative pathway, by using mesoscale simulations to quantitatively characterise the transient mesoscale dynamics of erythrocytes. The focus has been put on the quantitative description of the relaxation time of a single RBC following the sudden arrest of an external mechanical load. Our simulations featured a hybrid Immersed Boundary-Lattice Boltzmann Method (IB-LBM)~\cite{thesis:kruger}, further coupled with the Standard Linear Solid model (SLS)~\cite{art:lizhang19} to account for the RBC membrane viscosity. Two load mechanisms have been considered: RBC stretching by means of an external force applied at the ends of the erythrocyte (stretching experiment) and the hydrodynamic deformation induced by a given external shear (shear experiment). The relaxation time of the RBC membrane has been quantitatively characterised as a function of the magnitude of the mechanical load, while keeping the membrane viscosity as an input parameter. In the limit case of vanishing load strengths, the relaxation time is a linear function of the membrane viscosity and does not sensibly depend on the load mechanism: in such a limit, in fact, the intrinsic relaxation time of the membrane is recovered. For finite load strength magnitudes, however, the relaxation time proves to depend non trivially on the load strength. More specifically, considering realistic RBC relaxation times corresponding to a membrane viscosity in the range $\mum\in[1.59, 3.18]\times 10^{-7}$~m~Pa~s, results of numerical simulations show that for the stretching experiment the relaxation time decreases linearly with the load strength, while for the shear experiment it decreases with a non-linear behaviour. Overall, one could fairly conclude that while the order of magnitude of the relaxation time is set by the value of the membrane viscosity, its detailed functional dependency on the load strength is closely related to the typology of the load mechanism. Importantly, thanks to systematic investigations like those that we proposed in this study, such dependencies can be studied and quantified.\\
Hence the natural choice for a CFD model of VAD circulation would be that of a flow-dependent relaxation time. Our mesoscale simulations offer the possibility to ``tune'' such flow-dependency in a realistic way. Whether or not this spatial dependency may cause a sensible change in the predictions of the VAD performance definitively warrants future investigations.\\ 
More generally, it would be definitively of interest to study the time response of RBCs to a ``generic'' time dependent strain rate, which reproduces the main features of fluids of those encountered in biomedical devices. Following earlier studies on droplet dynamics~\cite{biferale2014deformation}, one can think of generating such signal from a dedicated numerical simulation, with the relevant dimensionless numbers matching the fluid dynamics of cardiovascular systems. One can then study the resulting RBC dynamical properties (deformation, accumulated stress, correlation with hemolysis) via the tensor-based models or even in LBM simulations themselves. The methodology here developed surely poses appropriate grounds for such an interesting future development.

\section*{Conflicts of interest}
There are no conflicts to declare.
\section*{Acknowledgments}
The authors acknowledge fruitful discussions with G. Koutsou, T. Kr\"{u}ger, S. Gekle, M. De Tullio, P. Decuzzi and P. Lenarda. This project has received funding from the European Union Horizon 2020 research and innovation programme under the Marie Sk\l odowska-Curie grant agreement No 765048. We also acknowledge support from the project ``Detailed Simulation Of Red blood Cell Dynamics accounting for membRane viscoElastic propertieS'' (SorCeReS, CUP N. E84I19002470005) financed by the University of Rome ``Tor Vergata'' (``Beyond Borders 2019'' call).


\renewcommand\refname{References}

\bibliography{main.bib} 
\bibliographystyle{rsc} 

\end{document}